\documentclass[]{aa}  
\usepackage{graphicx}
\usepackage{txfonts}
\usepackage{color}

\begin{document} 

\titlerunning{Quasar Ly$\alpha$ and PopIII diagnostics}
\authorrunning{Humphrey et al.}

   \title{Photoionization models for extreme Ly$\alpha$
     $\lambda$1216 and HeII $\lambda$1640 ratios in quasar halos, and
     PopIII vs AGN diagnostics}

   \subtitle{}

   \author{A. Humphrey\inst{1}, M. Villar-Mart\'{i}n\inst{2},
     L. Binette\inst{3}, R. Raj\inst{1}\fnmsep\inst{4}
                    }

   \institute{$^1$Instituto de Astrof\'{i}sica e Ci\^encias do Espa\c{c}o,
     Universidade do Porto, CAUP, Rua das Estrelas, PT4150-762 Porto, Portugal\\
              \email{andrew.humphrey@astro.up.pt}\\
 $^2$Centro de Astrobiolog\'{i}a (INTA-CSIC), Ctra de Torrej\'on a Ajalvir, km 4, E-28850 Torrej\'on de Ardoz, Madrid, Spain\\
 $^3$Instituto de Astronom\'{i}a, Universidad Nacional Aut\'onomo de M\'exico, Ap. 70-264, 04510 M\'exico D.F., M\'exico\\
 $^4$Indian Institute of Technology Guwahati, Guwahati 781039, India\\
}

   \date{Accepted for publication in A\&A on 8/10/2018}

 
  \abstract
   {}
   {We explore potential mechanisms to produce
     extremely high Ly$\alpha$/HeII flux
     ratios, or to enhance the observed
     number of Ly$\alpha$ photons per incident ionizing photon, in
     extended AGN-photoionized nebulae at high-redshift.}
   {We compute models to simulate, in the low density regime, 
     photoionization of interstellar gas by the radiation field
     of a luminous AGN. We explore the impact of
     ionization parameter, gas metallicity, ionizing spectrum,
     electron energy distribution, and cloud viewing angle on the
     relative fluxes of Ly$\alpha$, HeII and other lines, and on the
     observed number of
     Ly$\alpha$ photons per incident ionizing photon. We compare
     our model results with recent observations of quasar Ly$\alpha$
     halos at z$\sim$3.5.}
   {Low ionization parameter, a relatively soft or filtered
     ionizing spectrum, low gas metallicity, $\kappa$-distributed
     electron energies, or reflection of Ly$\alpha$ photons by neutral
     hydrogen can all result in significantly enhanced
     Ly$\alpha$ relative to other lines ($\ge$ 10 \%), with log
       Ly$\alpha$/HeII reaching values of up to 4.6. In the cases of
     low gas metallicity, reflection by HI, or a hard or
       filtered ionizing spectrum, the observed number of Ly$\alpha$ 
     photons per incident ionizing photon is itself significantly
     enhanced above the nominal Case B value of 0.66 due to
     collisional excitation, reaching values
    as high as 5.3 in an `extreme case' model which combines several
    of these effects. We find that at low gas metallicity
    (e.g. $Z/Z_{\odot}$=0.1) the production of Ly$\alpha$ photons is
    predominantly via collisional excitation rather than by
    recombination. In addition, we find that the collisional
    excitation of Ly$\alpha$ becomes much more efficient if the
    ionizing continuum spectrum has been pre-filtered through an
    optically thin screen of gas closer to the AGN (e.g., by a wide-angle,
    feedback-driven outflow.) We also show that the
Ly$\alpha$ and HeII emission line ratios of the z$\sim$3.5 quasars studied by
Borisova et al. (2016) are consistent with AGN-photoionization of gas
with moderate to low metallicity and/or low ionization parameter, without
requiring exotic ionization/excitation mechanisms such as strong
line-transfer effects. In addition, we present a set of UV-optical
diagnostic diagrams to distinguish between photoionization by Pop III
stars and photoionization by an AGN.}
   {}

   \keywords{}

   \maketitle

\section{Introduction}
The Lyman-alpha (Ly$\alpha$ $\lambda$1216) line of neutral hydrogen is 
an important observable in that it permits the discovery and study of
galaxies at high redshift (e.g. Shibuya et al. 2017a), particularly
galaxies in a phase of star formation or active galactic nucleus (AGN)
activity. This line is one of the intrinsically most luminous in
galaxy spectra, but is relatively unstable as a quantitative
diagnostic due to its resonant nature, and also due to its
susceptibility to extinction by dust or other effects (e.g. Binette et
al. 1993). Despite its significant disadvantages, Ly$\alpha$ remains a
key source of observational information about the evolution of
galaxies and their gaseous environments across cosmic time 
(e.g. Haiman and Rees 2001).

Although typically less luminous and thus more difficult to observe,
the non-resonant recombination line HeII $\lambda$1640 offers
additional information to that provided by Ly$\alpha$. Crucially, the
ionization potential of He$^+$ is far higher than that of H$^0$ (54.4
vs 13.6 eV) and the production of a significant flux of HeII relative
to Ly$\alpha$ requires an ionizing source with a relatively hard
ionizing spectrum, or else significant collisional ionization by
shocks.

Used together, Ly$\alpha$ and HeII offer potentially
powerful diagnostics of the nature of ionizing sources in high-z
galaxies, and a toolbox for understanding the physics of gaseous
nebulae therein (e.g. Heckman et al. 1991; Villar-Mart\'{i}n et
al. 2007; Arrigoni-Battaia et al. 2015a; Feltre, Charlot \& Gutkin
2016; Husemann et al. 2018). 

One potentially powerful application of the Ly$\alpha$ to HeII flux 
ratio is in the identification of Population III (Pop III) stars in
the high redshift Universe, by searching for HII regions or star
forming galaxies with an emission line spectrum indicative of
photoionization by an extremely hot thermal source (e.g., Schaerer et
al. 2002). However, care is
needed to avoid misclassifying AGN as Pop III and vice versa (see
e.g. Fosbury et al. 2003; Binette et al. 2003; Villar-Mart\'{i}n,
Cervi\~no \& Gonz\'alez Delgado 2004; Sobral et al. 2015, 2017; Bowler
et al. 2016).

In addition, the Ly$\alpha$ to HeII flux ratio can inform us about the
physics in nebulae associated with powerful AGN. For example, transfer
effects and dust can strongly affect this line ratio (e.g. van Ojik et
al. 1994; Villar-Mart\'{i}n, Binette \& Fosbury 1996), and
photoionization modelling also suggests that the gas density and
metallicity and the presence of young stars may also impact this line
ratio in nebulae that are photoionized by an AGN (Villar-Mart\'{i}n et
al. 2007).

\begin{figure*}
\includegraphics{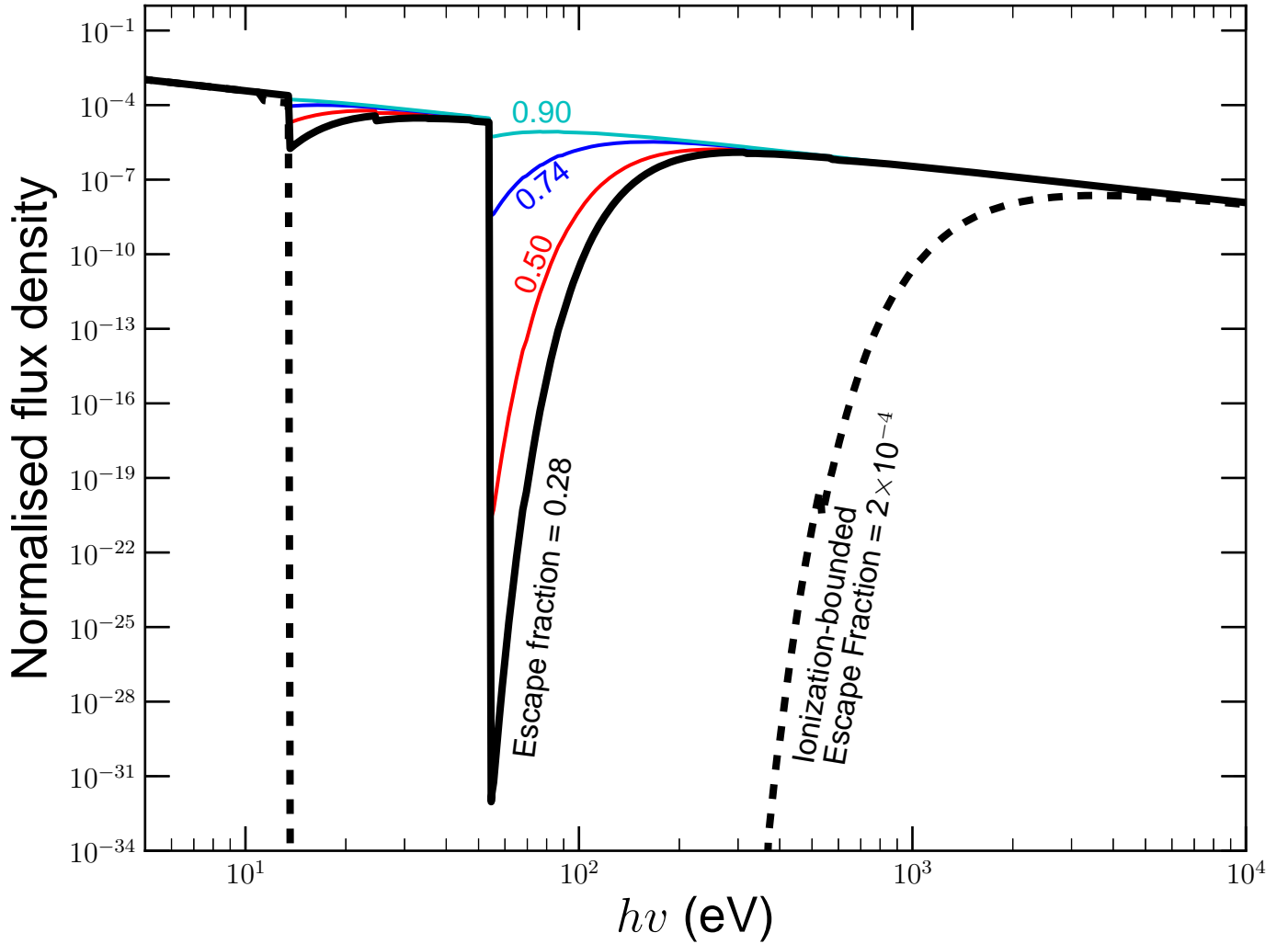}
\vspace{5.6in}

\vspace{0.0in}
\caption{Filtered AGN spectral energy distributions described
  in Sect. ~\ref{filtering}. These correspond to escape fractions of 0.28 
  (solid black line), 0.50 (solid red line), 0.74 (solid blue line)
  and 0.90 (solid cyan line). For comparison, we also show the
  emergent distribution from a typical ionization-bounded model
  (dashed black line): the UV part of the spectrum has been
  completely absorbed, while some X-rays still escape through the
  cloud. The vertical axis (normalised flux density or $S_{\nu}$)
  gives the power per unit area per unit frequency in arbitrary
  units. Note that the escape fractions are in photon number
    flux, while the SEDs are plotted in energy flux per unit frequency ($S_{\nu}$).}
\label{filtered}
\end{figure*}

Building on previous modeling (e.g., Villar-Mart\'{i}n, Binette \&
Fosbury 1996; Villar-Mart\'{i}n et al. 2007, Arrigoni-Battaia et
al. 2015a,b), we present photoionization model calculations appropriate
for large-scale Ly$\alpha$ nebulae photoionized by an AGN. In
particular, we explore several potential mechanisms to produce
`extreme' Ly$\alpha$ and HeII emission line ratios, or an enhancement
in the number of Ly$\alpha$ photons per incident ionizing photon. In 
addition, we explore possible diagnostic diagrams to distinguish
between photoionization by an AGN and photoionization by PopIII-like
stars. 

\section{Defining `extreme' Ly$\alpha$ flux ratios}
This study was motivated in part by the recent discovery by Borisova
et al. (2016) of Ly$\alpha$ halos around a number of quasars at
z$\sim$3.5, with Ly$\alpha$ $\lambda$1216 /HeII $\lambda$1640 flux
ratios that are typically significantly higher than those measured
from the spatially integrated spectra of powerful radio galaxies at
z$>2$ (e.g., Vernet et al. 2001; Villar-Mart\'{i}n et al. 2007). 

Following Villar-Mart\'{i}n et al. 2007, we
consider Ly$\alpha$/HeII $\lambda$1640 $>$ 15 (1.18 in log) to be an
`extreme' ratio. This is above the normal range of values shown by
high-z radio galaxies (see also De Breuck et al. 2000; Vernet et
al. 2001), and above the range of values produced by ionization-bounded
photoionization models that are able to reproduce the UV line ratios
of z$\sim$2 radio galaxies (see, e.g., Villar-Mart\'{i}n, Tadhunter \&
Clark 1997, Villar-Mart\'{i}n et al. 2007). 

\section{Photoionization modeling}
In order to examine how various physical conditions and the nature of
the ionizing source can affect the emergent emission line spectrum of
a photoionized nebula, we have computed a grid of photoionization
models using the multi-purpose modeling code MAPPINGS 1e
(Binette, Dopita \& Tuohy 1985; Ferruit et al. 1997; Binette et
al. 2012). 

Some regions of parameter space we investigate here have
also been explored previously (e.g., Villar-Mart\'{i}n,
Binette \& Fosbury 1996; Villar-Mart\'{i}n et al. 2007;
Arrigoni-Battaia et al. 2015a,b). Here we build on existing work by
considering a wider range of parameter space, additional
parameters, and previously unexplored combinations of parameters. 

Note that 2-photon emission is present in all our models, due to
  the modeled plasma being in the low-density regime.

\subsection{Fixed Parameters}
We keep several model parameters fixed throughout this study. 
For simplicity, we adopt an isochoric, plane-parallel, single-slab
geometry. By default, the models are ionization-bounded unless
otherwise stated, with computation terminated when the ionization
fraction of hydrogen falls below 0.01.

\subsection{Gas Density $n_H$}

Throughout this paper, we define $n_H$ as the Hydrogen gas number
density within our line emission models. Thus, all values of $n_H$ discussed
herein are not necessarily directly equivalent to the {\it volume-averaged}
gas densities which are often preferred when modeling large scale
structure (e.g., Rosdahl \& Blaizot 2012). 

We have adopted 100 cm$^{-3}$ as our
default hydrogen number density,  because we aim to simulate spatially
extended, low density gas rather than compact regions of high density
gas closely associated with the central AGN, whose gas clouds typically have
$n_H$$>$1000 cm$^{-3}$. To test the impact of lower gas density on the
Ly$\alpha$ emission, we have computed additional models using
$n_H$=0.1 cm$^{-3}$, corresponding to the volume-averaged gas density
in the outskirts of high-z quasar halos predicted by the numerical
simulations of Rosdahl \& Blaizot (2012).

\subsection{Ionization Parameter U}

The ionization parameter\footnote{$U=\frac{Q}{4 \pi r^2 c 
    n_H}$, where $Q$ is the luminosity 
of ionizing photons emitted by the ionizing source assuming it is isotropic, $r$ is the
distance of the cloud from the ionizing source, and $n_H$ is the
hydrogen density of the cloud.} U is effectively a measure of the ratio
of ionizing photons to particles in the model gas cloud, and provides a
useful parameterization of the ionization state of the gas. High-z
Ly$\alpha$ emitters can be very large ($r>$100 kpc; ), and 
there is a potentially very large range in U between individual objects and
between different regions of objects, depending on $r$, $Q$, $n_H$, etc. For
example, a cloud of gas in the outer Ly$\alpha$ halo of a luminous
quasar might have log $U \sim$ -3.5 (assuming $Q\sim$$10^{56}$
s$^{-1}$, $r\sim$100 kpc, $n_{H}\sim$10 cm$^{-3}$). A cloud 
closer to the quasar for instance might have log $U \sim$ -1.5 (with $r\sim$1 kpc,
$n_H$$\sim$1000 cm$^{-3}$). On the other hand, a cloud might have
log $U \sim$ -4.5 with $n_H$$\sim$1 cm$^{-3}$ at a distance of
$r\sim$1 Mpc from a quasar. Thus, we should consider a suitably large 
range in U. Our model grid contains 100 values of U, starting at
log U = -5 and ending at log U = 0.25. We increase U by +0.0531 dex
(i.e., a factor of 1.13) between each model in the U-sequence to
obtain a relatively fine sampling of this parameter.

\begin{figure*}

\includegraphics{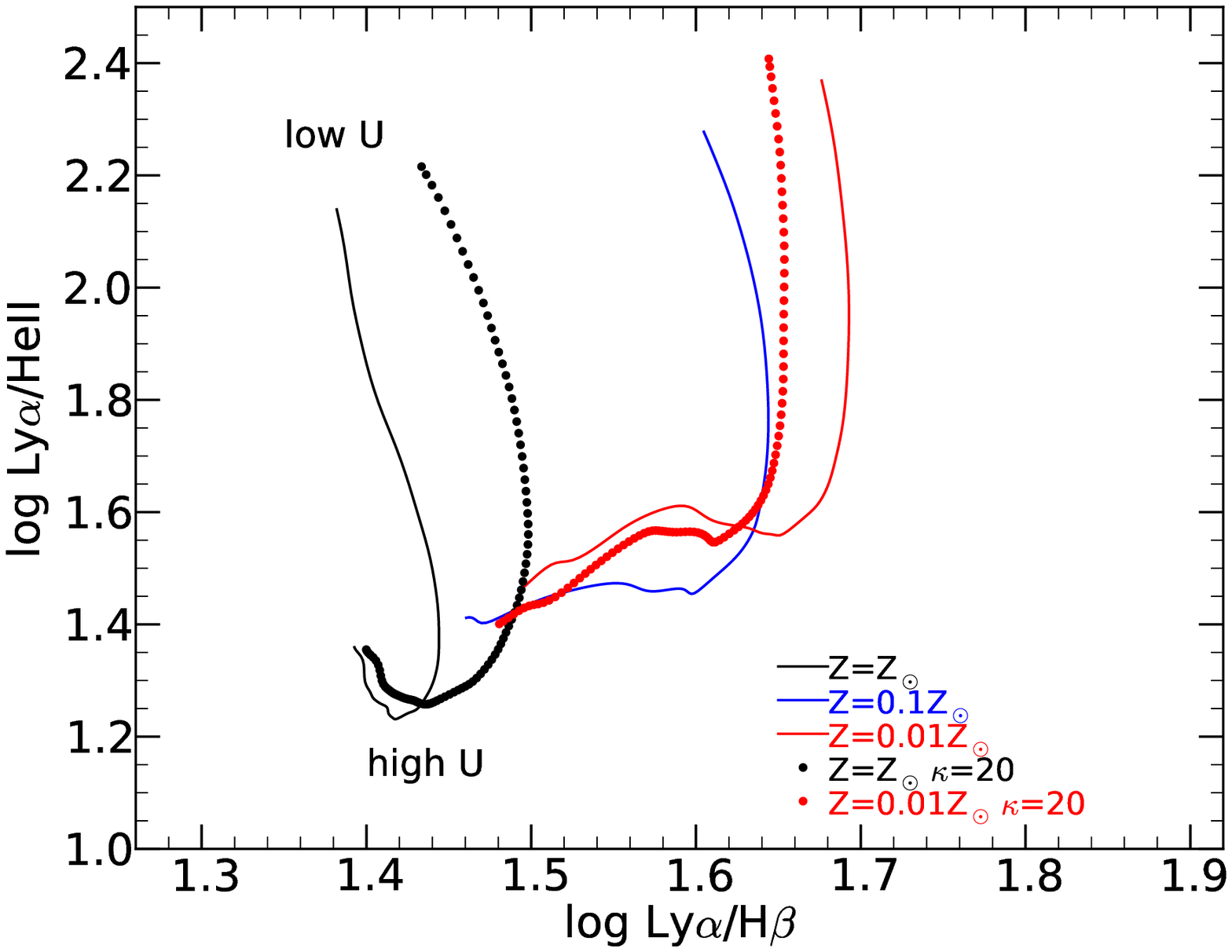}
\includegraphics{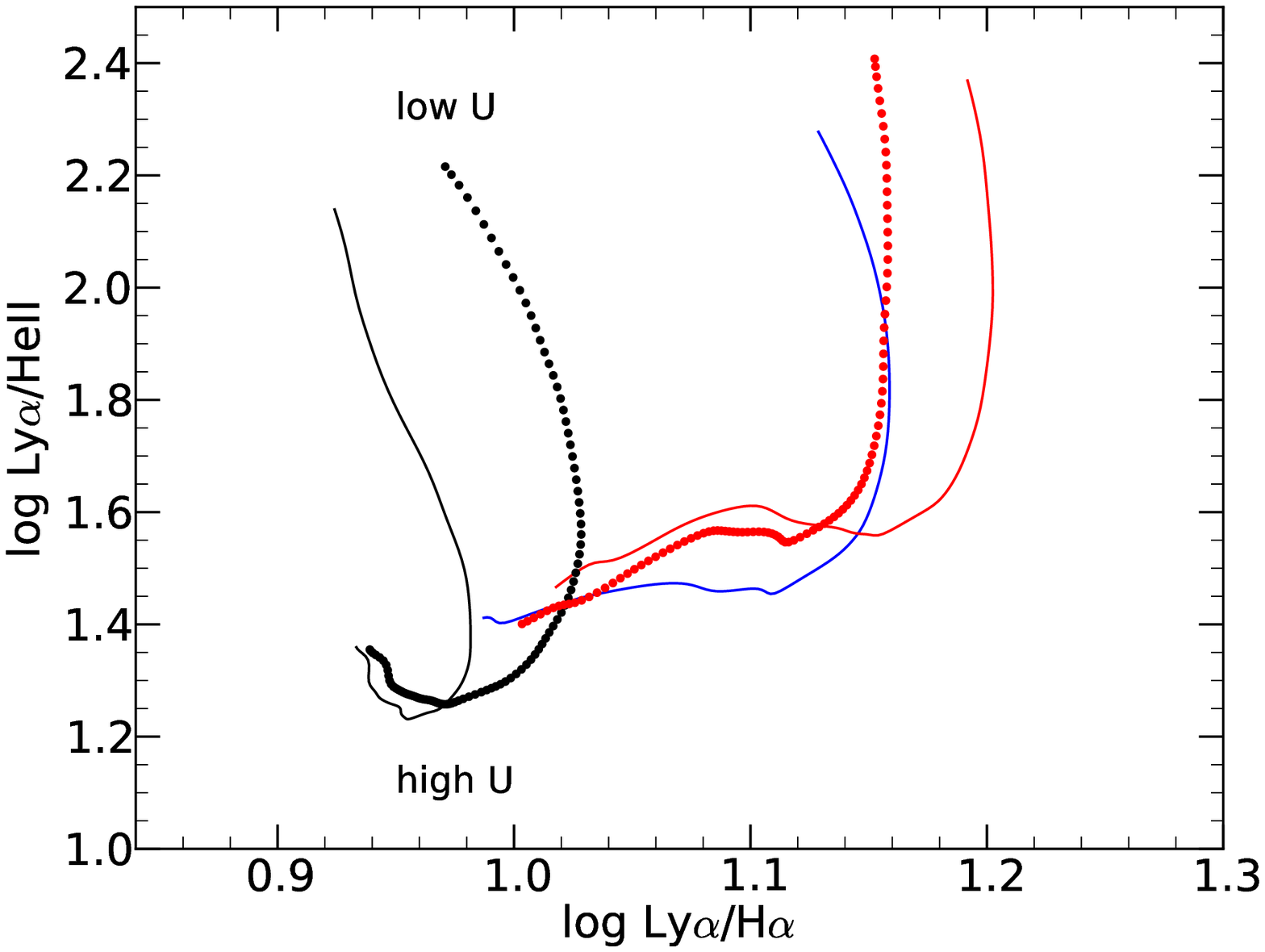}
\includegraphics{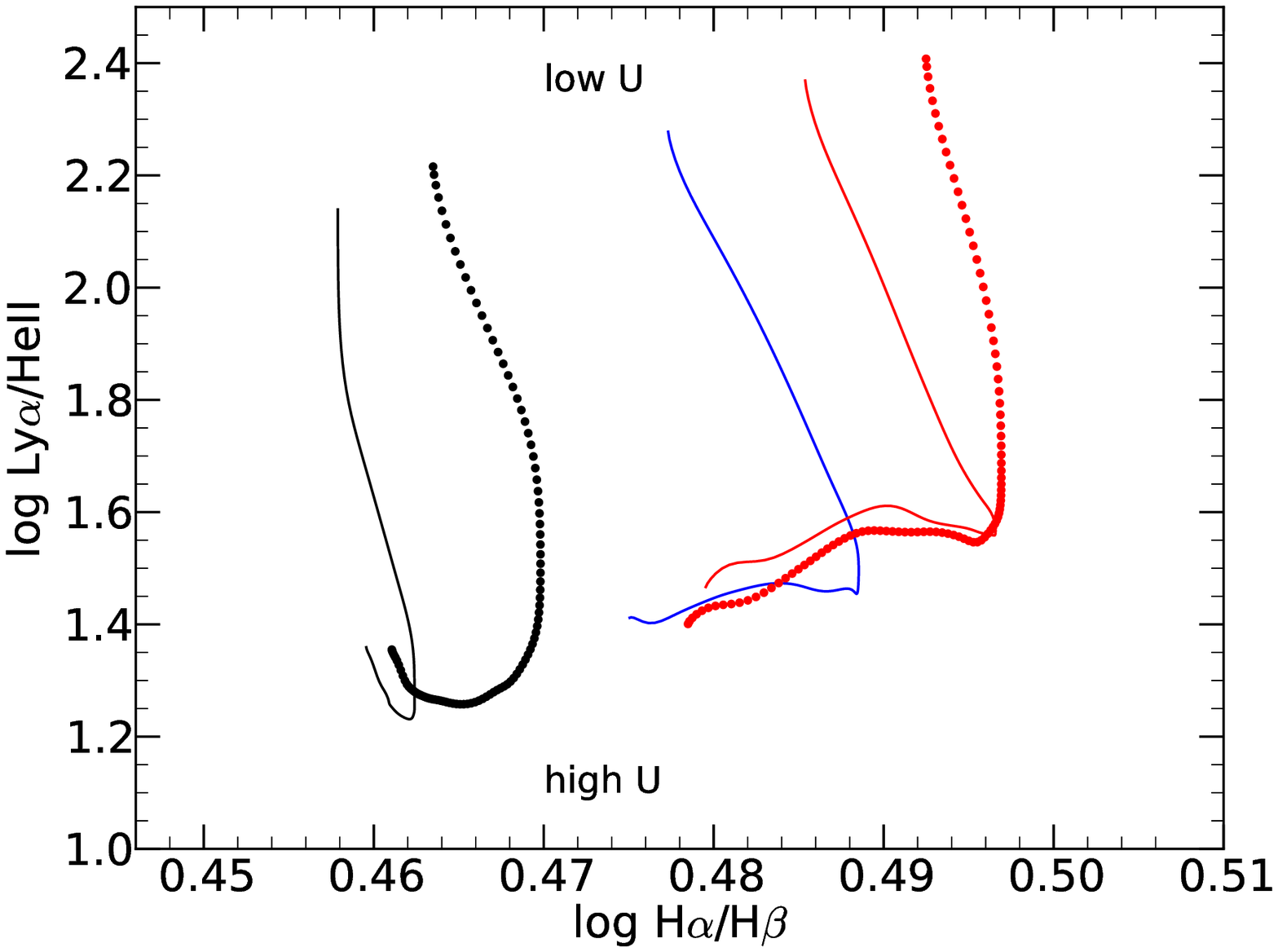}
\includegraphics{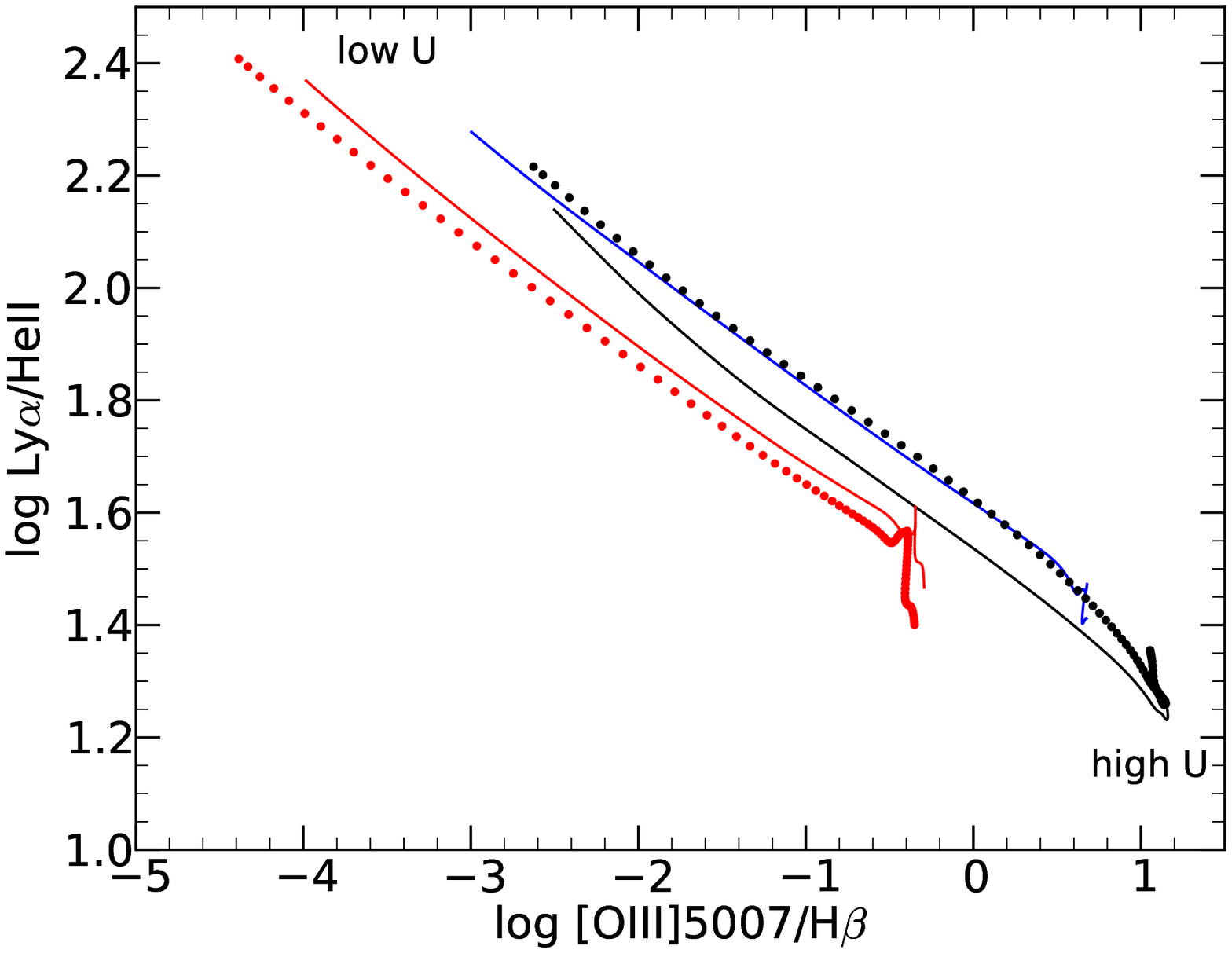}
\includegraphics{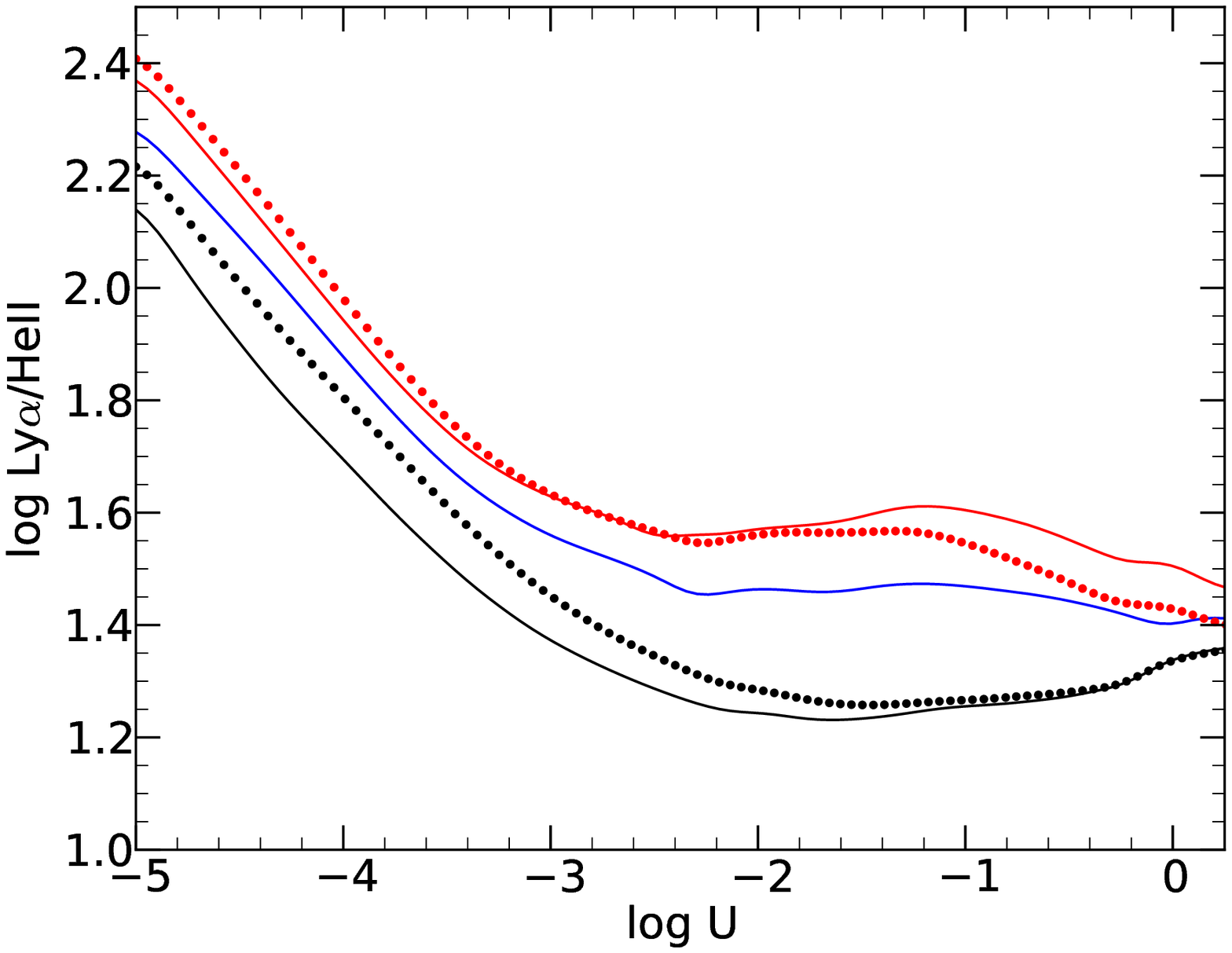}
\includegraphics{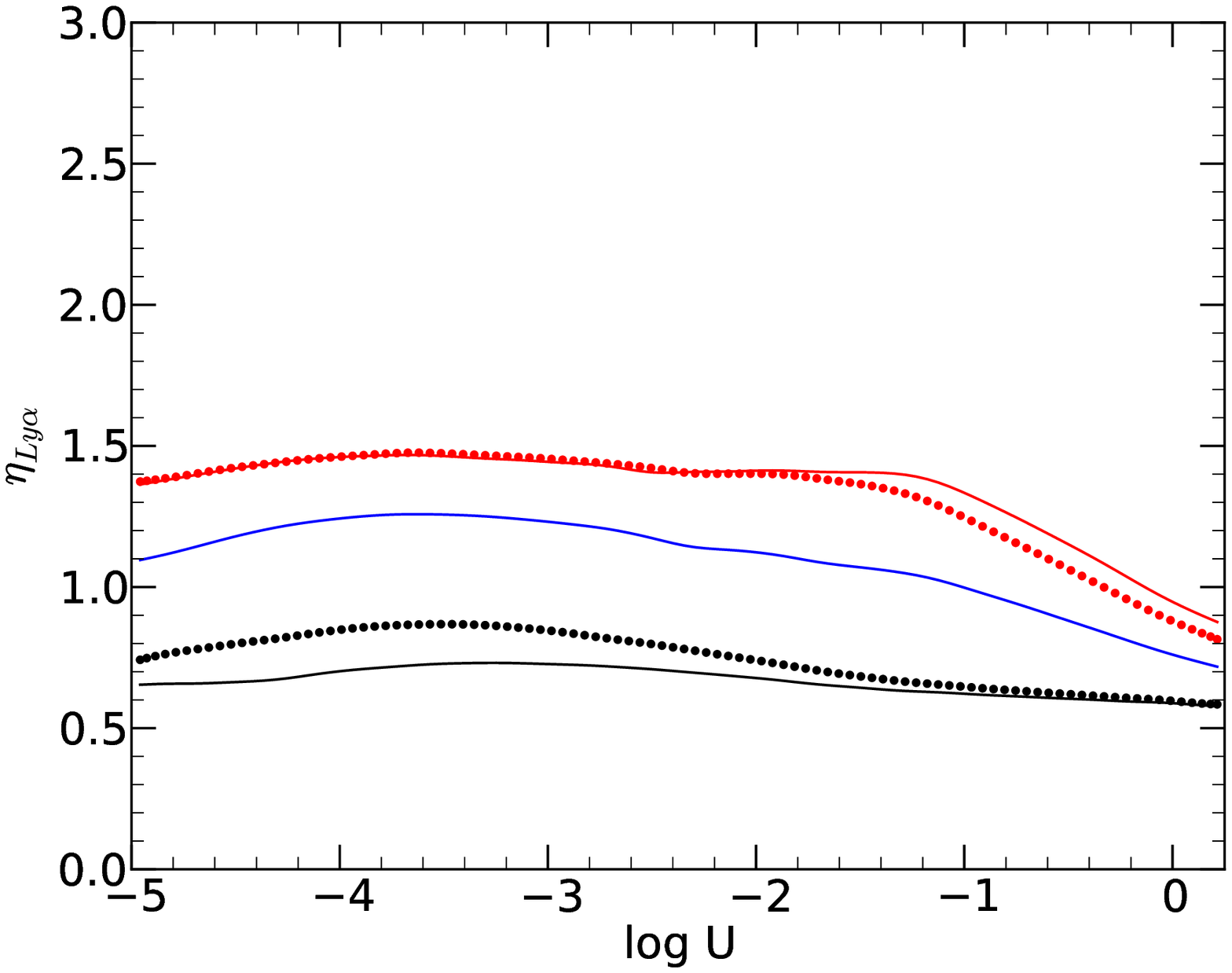}
\vspace{9.in}
\caption{Selected sequences in U plotted for Ly$\alpha$/HeII vs. Ly$\alpha$/H$\beta$,
  Ly$\alpha$/H$\alpha$, H$\alpha$/H$\beta$, or [OIII] $\lambda$5007 /
  H$\beta$. Each sequence represents a progression in U, for each of the three values of
  $Z$. Also shown are sequences in U using $\kappa$=20 for
  $Z/Z_{\odot}$=0.01 and 1.0 (dotted curves). All models in this
  figure use power index $\alpha=$-1.5. The lower two panels show
  Ly$\alpha$/HeII vs. U (left) and the ratio of Ly$\alpha$ photons to
  ionizing photons ($\eta_{Ly\alpha}$; right) for the same model
  sequences. The model loci cover the range of ionization
  parameter -5$<$log U$<$0.25.} 
\label{lya_balmer}
\end{figure*}

\subsection{Ionizing Powerlaw Index}
By default, we adopt a power law with the
canonical spectral index $\alpha$=-1.5 (e.g., Robinson et al. 1987),
where $S_{\nu}$ $\propto$ $\nu^{+\alpha}$ (see also Telfer et
al. 2002). However, it has been
suggested that AGN at high redshift (i.e., z$\ga$2) have a
significantly harder ionizing spectrum (e.g., Francis 1993; Villar-Mart\'{i}n
et al. 1997). On the other hand, it is conceivable that some active
galaxies at high redshifts may instead have softer ionizing SEDs, due
to the presence of a massive starburst which could contribute photons
to the ionization of the extended Ly$\alpha$ halo
(e.g. Villar-Mart\'{i}n et al. 2007), or perhaps due to partial
filtering of the AGN's ionizing continuum in a circumnuclear screen of
gas (e.g., Binette et al., 2003). To test the impact of adopting a harder or softer
powerlaw, we have also computed model sequences using $\alpha$=-1.0 or
$\alpha$=-2.0. 

\subsection{Gas Chemical Abundances}
We also vary the gas chemical abundances, to examine the impact of
metallicity on the emergent emission line spectrum. As the starting
point for the chemical abundance sets considered 
here, we adopt the Solar chemical abundances as determined by
Asplund et al. (2006). For non-Solar gas abundances, we scale all
metals linearly. 

Ly$\alpha$ emitting regions associated with galaxies at high-z have a
potentially very large dispersion in gas metallicity, ranging from
extremely metal poor gas on the outskirts of a very young galaxy at
high redshift, to metal-enriched gas close to the active nucleus of a
massive galaxy. Thus, we consider three representative gas
metallicities of $Z=$ 0.01, 0.10, and 1.0 times the Solar metallicity
($Z_{\odot}$). 

\subsection{Electron Energy Distribution and $\kappa$}
Among the novel features of MAPPINGS 1e is the ability to use the
non-equilibrium, Kappa-distribution (KD) of electron energies instead of the
commonly adopted Maxwell-Boltzmann distribution (MBD). The KD of
electron energies used in MAPPINGS 1e is a function of electron 
temperature and the $\kappa$ parameter, where 1.5 $<$ $\kappa$
$<\infty$, and is described in Nicholls et al. (2012; see also
Vasyliunas 1968). When $\kappa$ $\rightarrow$ $\infty$, the
distribution becomes a MBD. Full details of the implementation of KD
electron energies in MAPPINGS 1e are given in Binette et al. (2012). 

Although the physics underpinning the KD is not yet fully
understood (see, e.g., Ferland et al. 2016; Draine \& Kreisch 2018),
the presence of KD energies in Solar System plasmas is well
established (e.g. Vasyliunas 1968; Livadiotis \& McComas 2011;
Livadiotis 2018). Thus, it is important to consider the
possibility that KD electron energies may be present in some
extrasolar and extragalactic plasmas.

Indeed, a number of recent studies have shown that many commonly
observed nebular emission lines can be significantly affected by
adopting the KD, both in HII regions and high-z radio galaxies, and
the KD has been proposed as a solution to the long-standing
observational temperature discrepancies in planetary nebulae and HII
regions (Nicholls et al. 2012, 2013; Binette et al. 2012; Humphrey \&
Binette 2014). 

Values in the range 10$\la$$\kappa$$\la$40 have been proposed for HII
regions and planetary nebulae (e.g. Nicholls et al. 2012; Binette et
al. 2012) and AGN (Humphrey \& Binette 2014), where a smaller $\kappa$
represents a stronger deviation from the MBD. In this
Paper we consider $\kappa=$20 in place of the MBD distribution in some
model sequences. 

\subsection{Cloud Viewing Angle}
As previously shown by Villar-Mart\'{i}n,
Binette \& Fosbury (1996), neutral hydrogen at the back of the cloud
can act as a `mirror' to back-scatter Ly$\alpha$ towards the front of
the cloud, enhancing the flux of Ly$\alpha$ as seen by the observer,
relative to other lines. To avoid confusion with other
  geometrical set-ups, we refer to this neutral `mirror' at the rear of
  the cloud as a `back-mirror'.

We consider two viewing angles for the model cloud (slab). By default,
we have adopted the `side view', which places the observer at an angle
of 90$^{\circ}$ to the direction of travel of the incident ionizing 
radiation. In addition, we also calculate the `front view', where the
observer views the illuminated face of the cloud (see Binette et
al. 1993a,b).

In all of our ionization-bounded models, the column density of neutral
hydrogen towards the back of the cloud ($N_{HI}$ $\ge$ 10$^{18}$ cm$^{-2}$) is
substantially higher than what is required to produce a significant neutral
`back-mirror' effect (the minimum required is $N_{HI}$ $\ga$ 10$^{14}$ cm$^{-2}$). 

This `back-mirror' effect would be much weaker in
models in which the illuminated gas is optically thin to ionizing
radiation (matter-bounded), such as the filtering screen
described in Sect. ~\ref{filtering} below, or
the optically thin photoionization models that Arrigoni Battaia et
al. (2015b) favoured in explaining the Ly$\alpha$ emission from the
giant gas halo of the z=2.28 quasar UM287. 

Note that to reflect Ly$\alpha$ photons from the active nucleus
(whether hidden or viewed directly), such neutral `back-mirrors' as
considered here are not strictly needed. This is because even a
matter-bounded HII slab of reasonable opacity (e.g. Binette, Wilson \&
Storchi-Bergmann 1996) will have a non-negligible HI fraction and thus
can scatter to our line-of-sight Ly$\alpha$ photons from the active
nucleus (e.g. Humphrey et al. 2013a; Cantalupo et al. 2014),
potentially providing a further enhancement to the flux of
Ly$\alpha$. However, we do not consider this additional effect here.

\subsection{Ionization by a filtered SED}
\label{filtering}
We also study the impact of photoionizing gas clouds using an ionizing
powerlaw SED that has first been filtered through a screen of gas that
does not fully absorb all ionizing photons. A situation where this
might occur would be in a galaxy which contains a wind-blown
superbubble that is expanding into the extended gaseous 
halo (e.g., Tenorio-Tagle et al. 1999; Taniguchi et al. 2001). In
this case, one would expect the bubble to partially filter the AGN's
ionizing continuum so that gas clouds at radii beyond the bubble would
see an altered version of the ionizing SED. Alternatively, the
ionizing spectrum of the AGN might be filtered by gas in the host
galaxy, before escaping to ionize gas on hundreds of kpc scales,
giving essentially the same result. 

Several previous studies have addressed this issue (see e.g. Binette,
Wilson \& Storchi-Bergmann 1996; Binette et al. 2003), and among the
main effects are lower fluxes for HeII and other high-ionization
lines, and lower electron temperatures, than would have resulted using
the original (unfiltered) SED. 

Here, we look specifically at the impact of a filtered ionizing SED on
Ly$\alpha$ flux ratios. For this, we have produced a set of four
filtered SEDs by computing matter-bounded photoionization models with
$\alpha$=-1.5, log U=-2 and N$_H$ = 1.0, 2.5, 5.0 or 7.0
$\times$10$^{20}$ cm$^{-2}$, giving ionizing continuum escape
fractions of $F_{esc}$ = 0.90, 0.74, 0.50 and 0.28,
respectively\footnote{We define $F_{esc}$ as the fraction of H-ionizing
    photons that pass through the filtering screen of gas unabsorbed.}. The choice of
values for U and N$_H$ are not critical, and can be scaled in lockstep
to produce a similar emergent distribution. The metallicity of the
screening gas is assumed to be solar. The resulting
ionizing SEDs were then used as inputs for new, U-sequence model
calculations. 

The various filtered SEDs are shown in Fig. ~\ref{filtered}. The main
dips seen in our filtered SEDs are caused by 
photoelectric absorption of photons by H (13.6 eV) and He$^+$ (54.4
eV), resulting in a spectrum with a comparatively lower number of He$^+$
ionizing photons. For comparison, we also show the residual of an $\alpha$=-1.5
ionizing SED from an unfiltered model calculation. 

{When computing models using a filtered SED, we
  have scaled by $F_{esc}$ the range in U values covered. For
  instance, when using our $F_{esc}$=0.28 SED, our U-sequences run
  from log U = -4.55 to log U=-0.30. This introduces a differential
  shift in the curves that are plotted as a function of U. To
  facilitate visual comparison between models, we show in Appendix A
  the same models as a function of U$_{\ast}$ = U / $F_{esc}$.

Note that we do not include line emission produced within the
filtering screen in the resulting line fluxes from our filtered
continuum models. The relative significance of the line flux from the
absorbing screen depends strongly on its physical conditions, and can
varies from significant when the screen and post-screen gases
have similar physical conditions (see the models of Binette, Wilson \&
Storchi-Bergmann 1996), to negligible as in the case where the
filtering screen is obscured (from the observer) by nuclear dust. A
full treatment of the emission from such a screen is beyond the scope
of this Paper.

\begin{figure}
\includegraphics{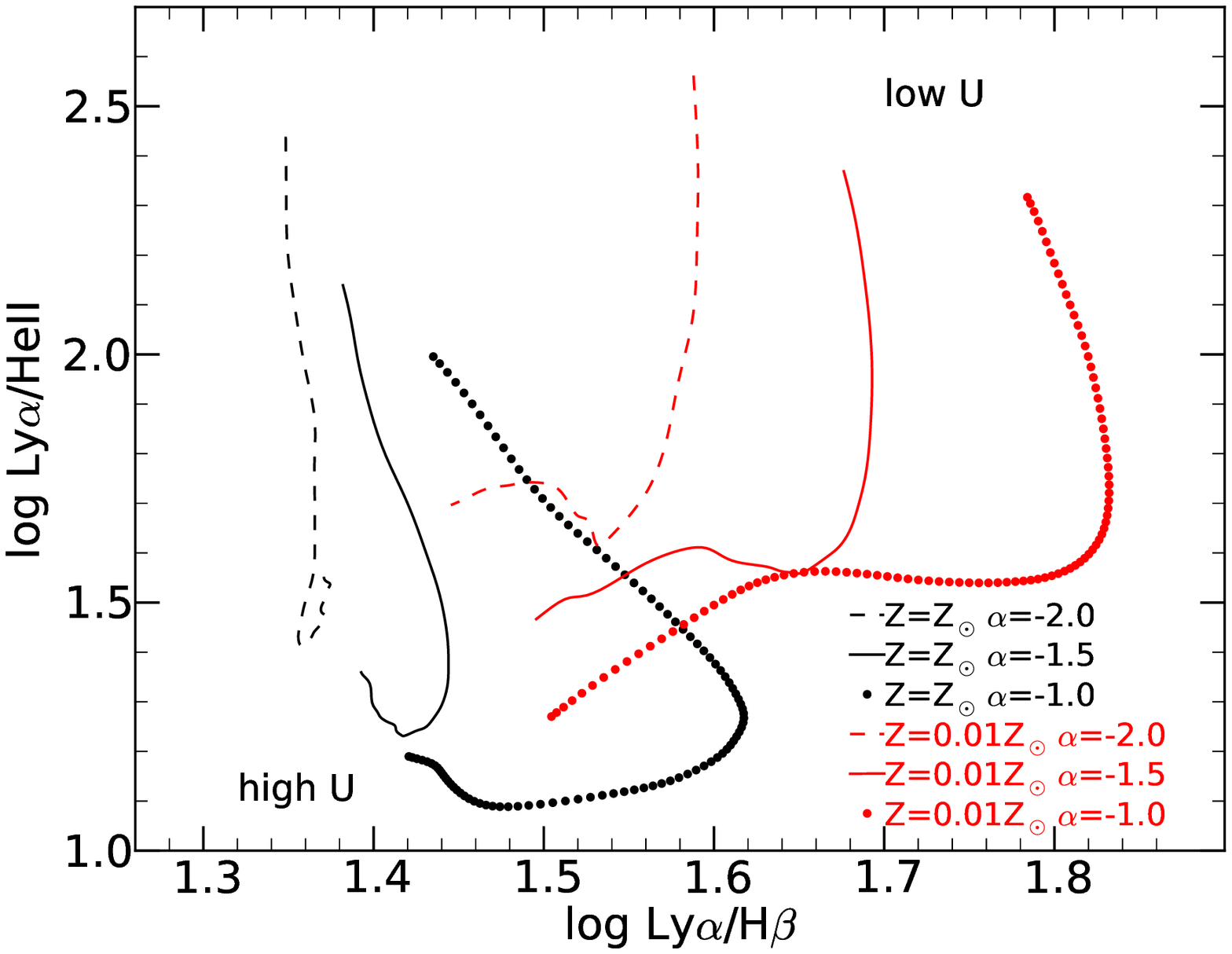}
\includegraphics{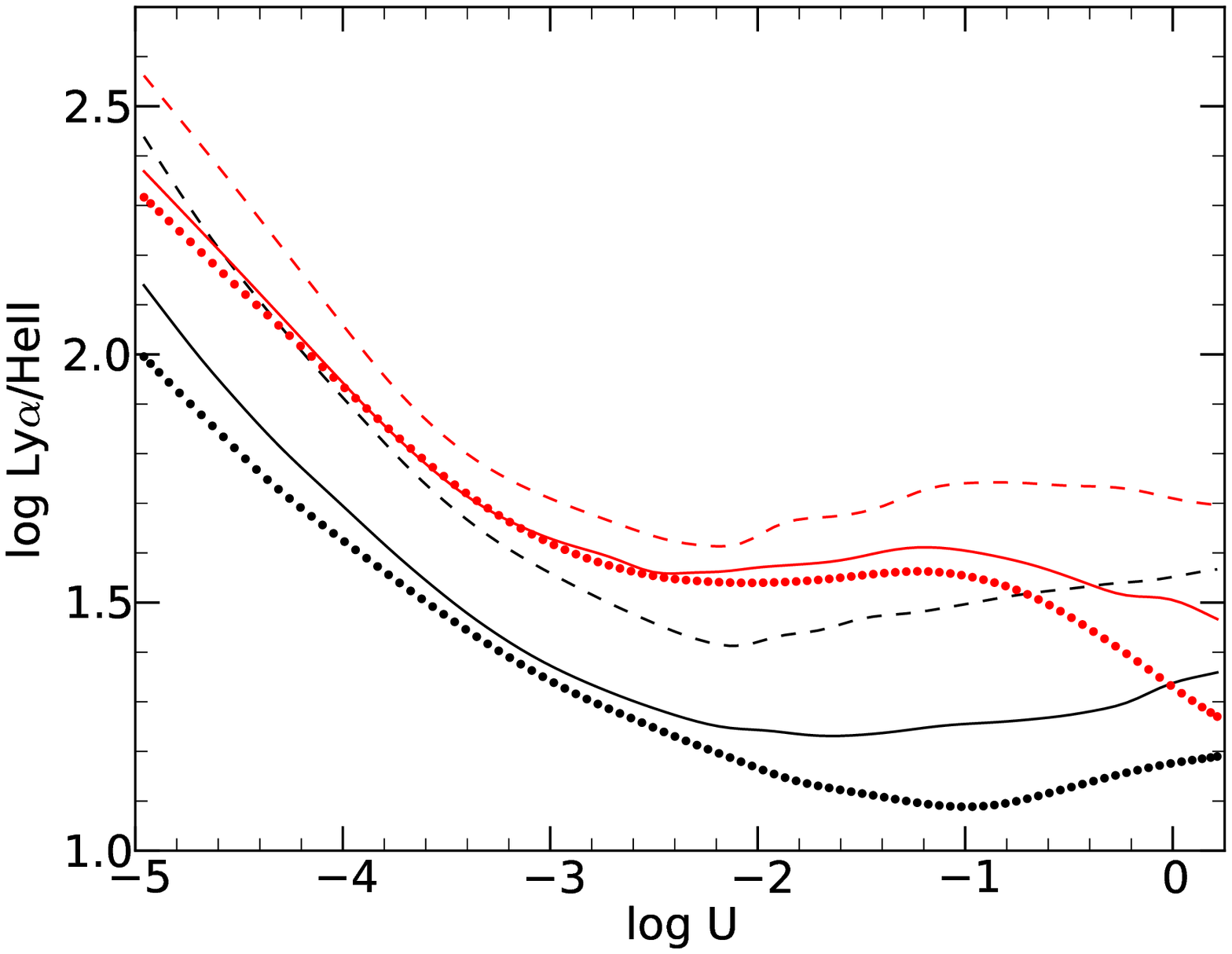}
\includegraphics{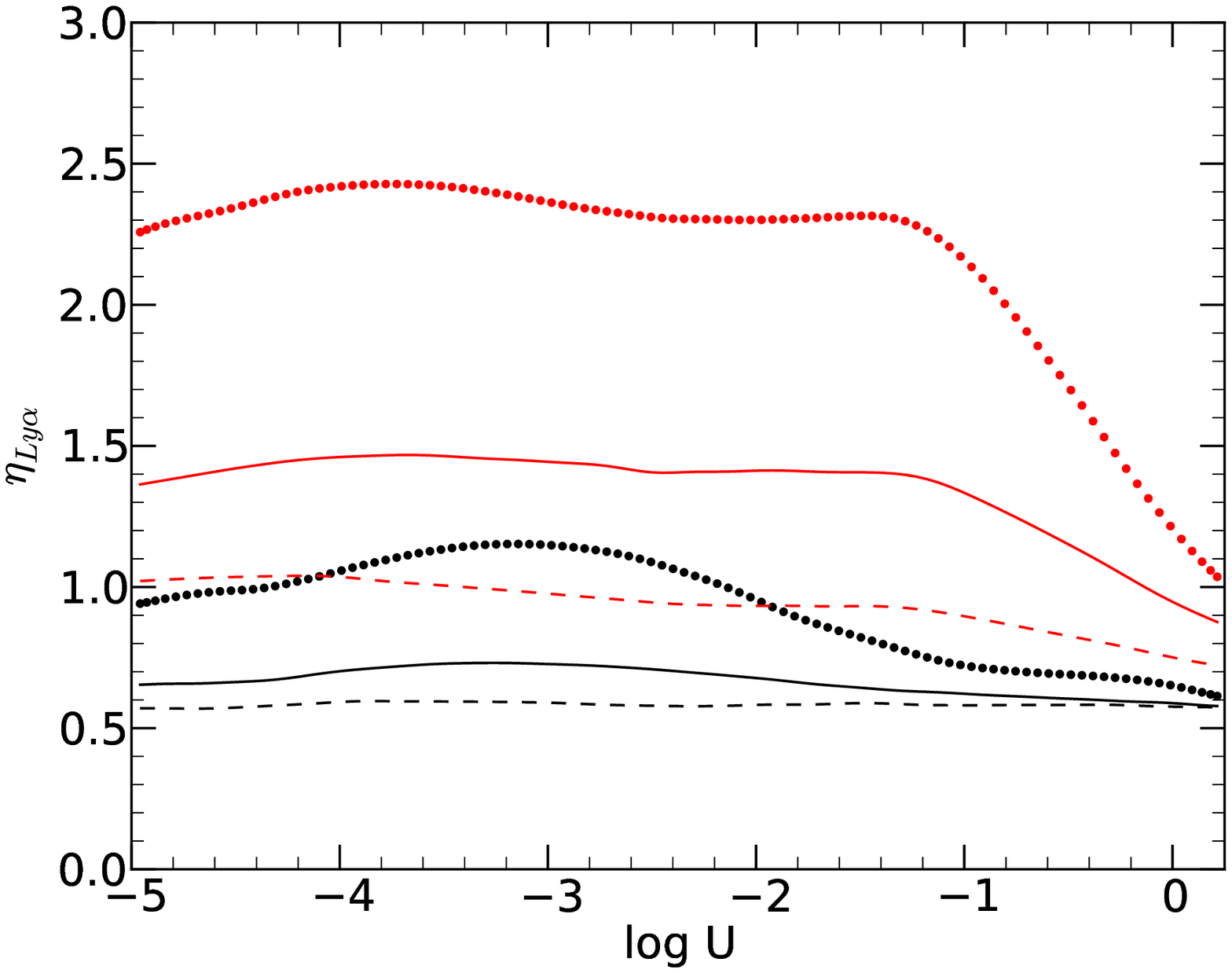}
\vspace{9in}
\caption{The impact of the powerlaw index of the ionizing spectrum on
  Ly$\alpha$/HeII, Ly$\alpha$/H$\beta$ and $\eta_{Ly\alpha}$. The model loci cover the range of ionization
  parameter -5$<$log U$<$0.25.}  
\label{alpha}
\end{figure}

\begin{figure}
\includegraphics{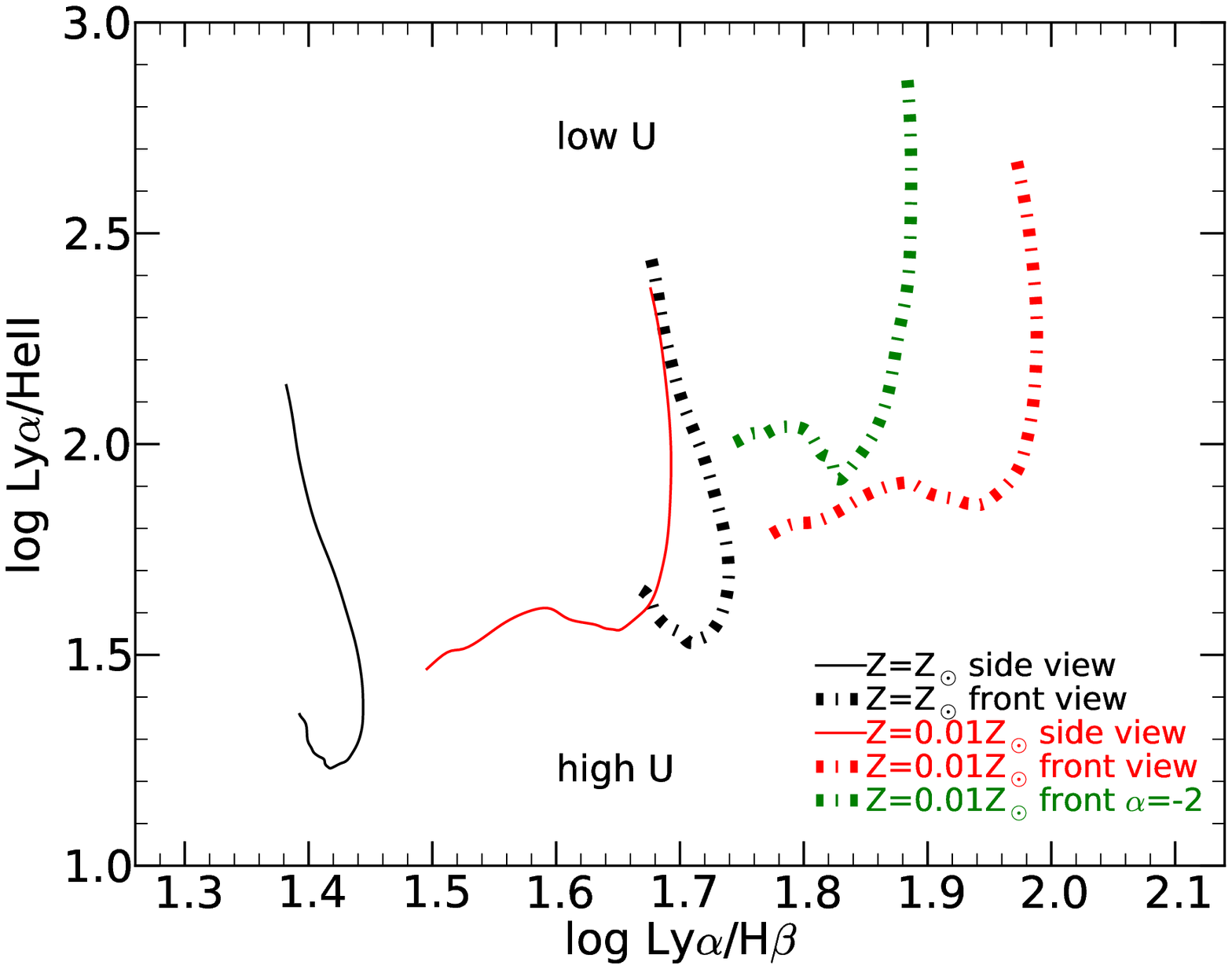}
\includegraphics{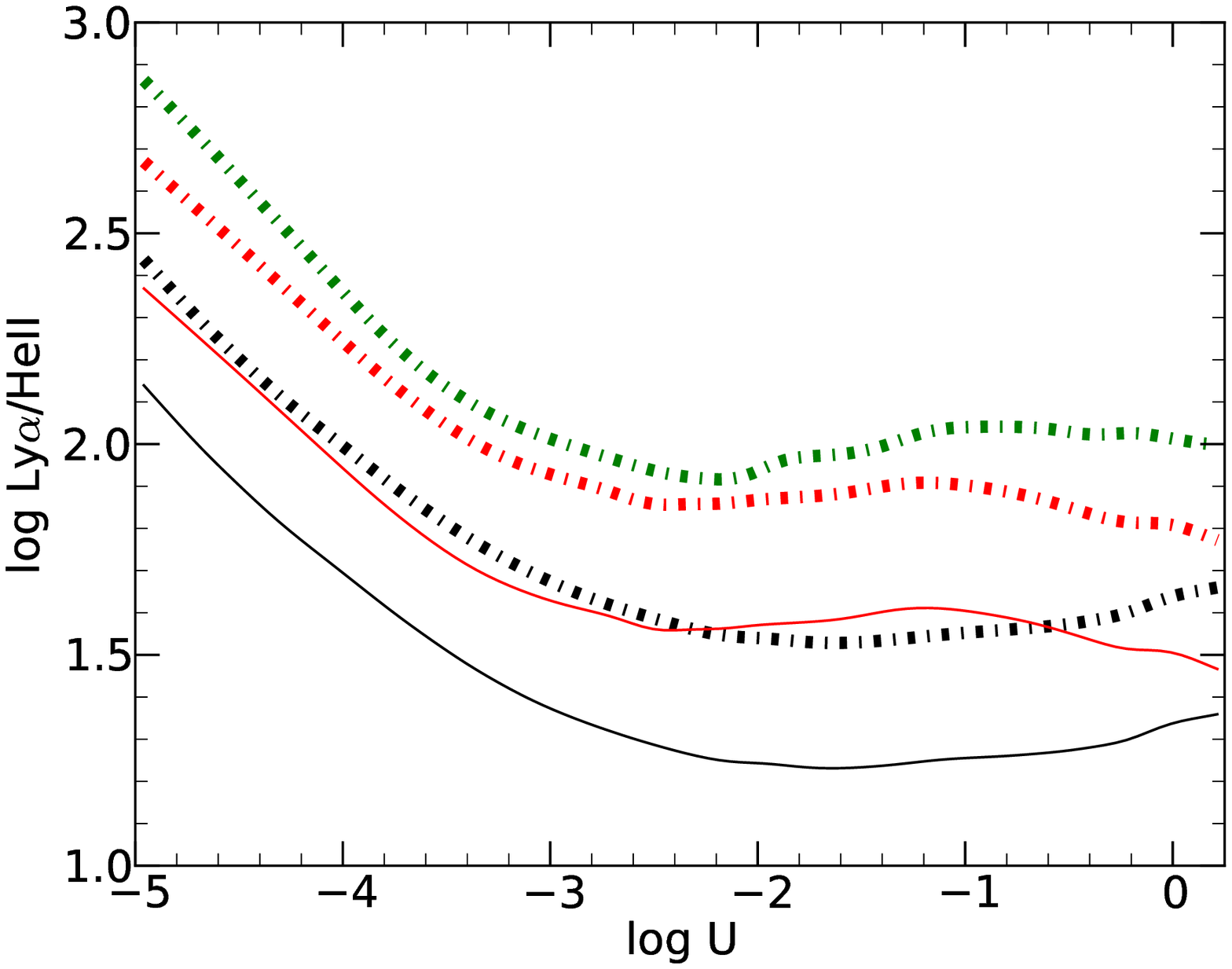}
\includegraphics{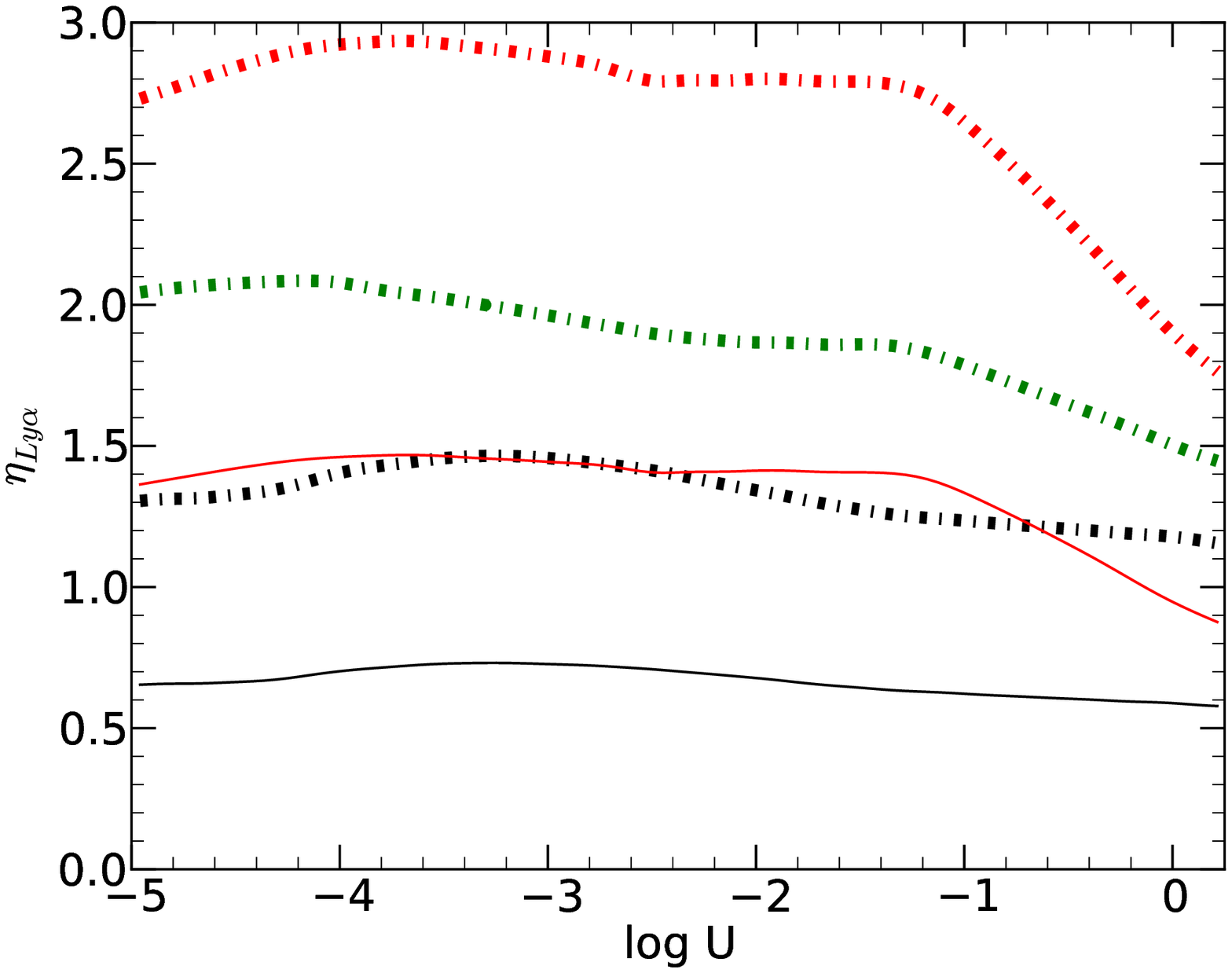}
\vspace{9in}
\caption{The impact of viewing angle on the observed Ly$\alpha$/HeII,
  Ly$\alpha$/H$\beta$ and $\eta_{Ly\alpha}$ values. The green curve
  shows the locus of our sequence in U using $Z$/$Z_{\odot}$=0.01,
  $\alpha$=-2.0 and the `front view', to illustrate the combined effect of
  low U, low gas metallicity, a relatively soft ionizing continuum,
  and a `back-mirror'. The model loci cover the range of ionization
  parameter -5$<$log U$<$0.25.}  
\label{mirror}
\end{figure}

\begin{figure}
\includegraphics{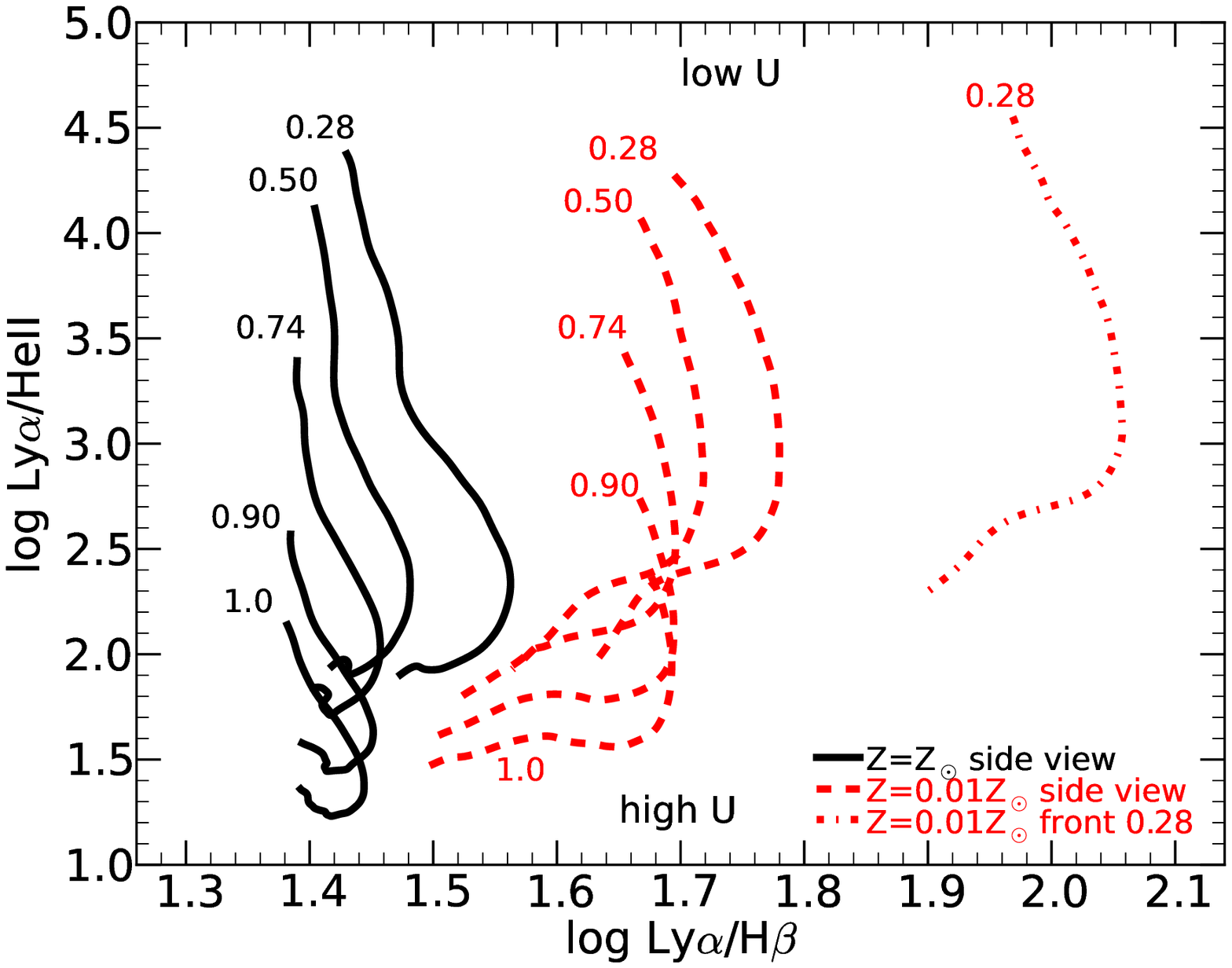}
\includegraphics{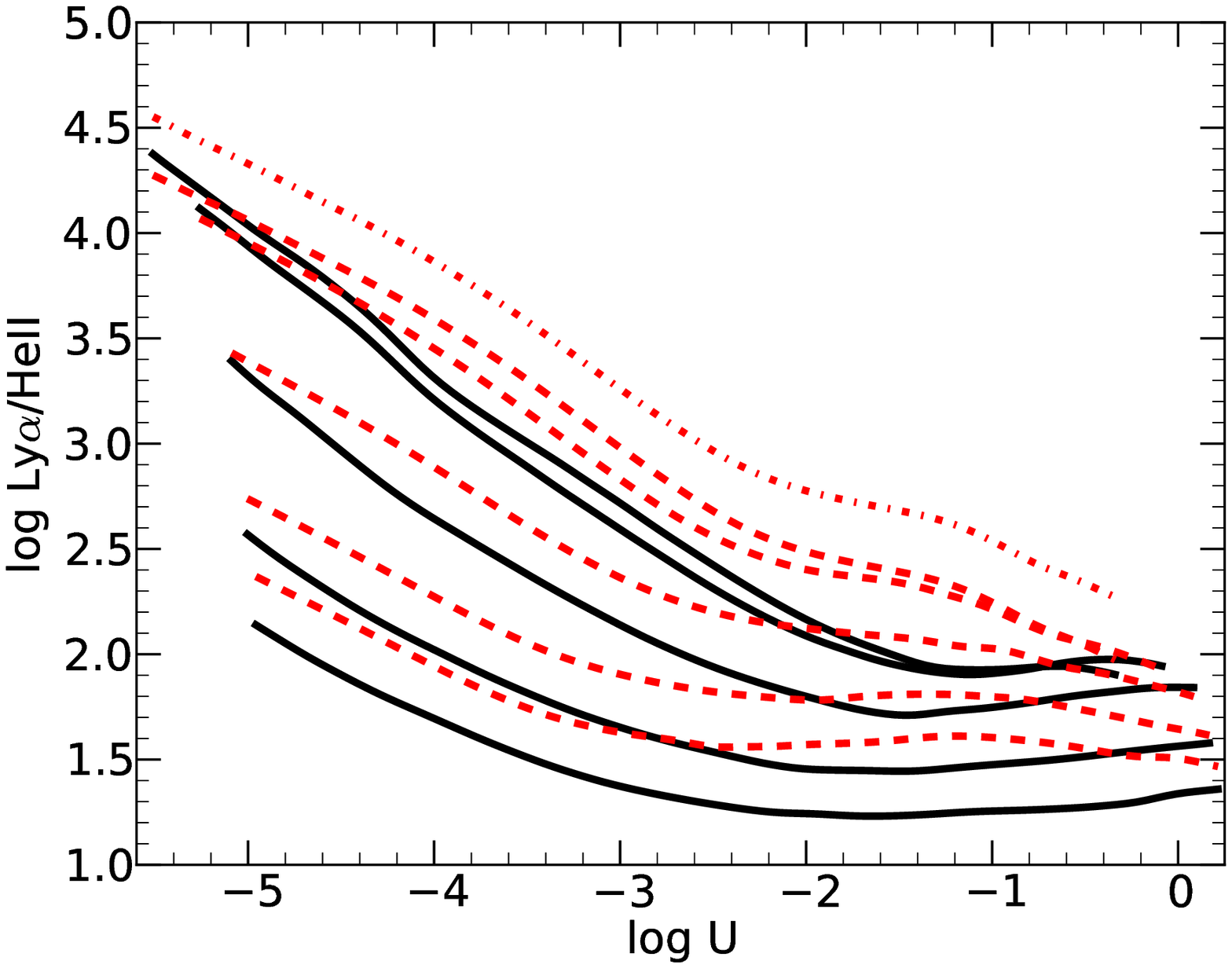}
\includegraphics{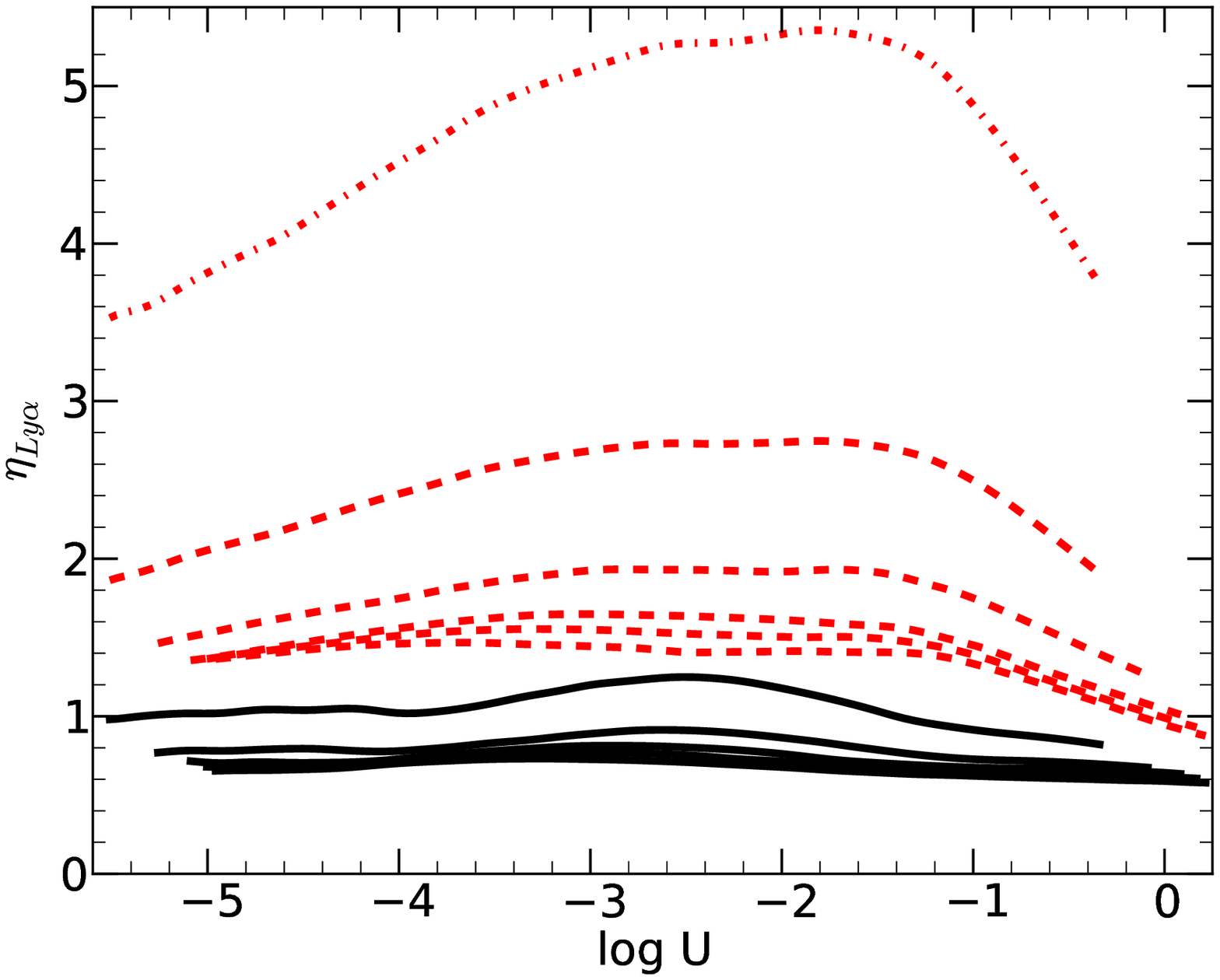}
\vspace{9in}
\caption{The impact of using a filtered ionizing continuum instead of
  a simple powerlaw on the observed values of Ly$\alpha$/HeII,
  Ly$\alpha$/H$\beta$ and $\eta_{Ly\alpha}$. U-sequences using a
  filtered ionizing continuum are labeled with their value of
  $F_{esc}$. U-sequence loci which use our default powerlaw of
  $\alpha$=-1.5 are labeled '1.0' because the input SED is
  unfiltered. The dot-dashed curve (on the right of the upper panel)
  shows the locus of our sequence in U that uses $F_{esc}$=0.28,
  $Z$/$Z_{\odot}$=0.01, $\alpha$=-1.5 and the `front view', to 
  illustrate the combined effect of low U, low gas metallicity, a heavily
  filtered ionizing continuum, and a `back-mirror'. For
each combination of U and $Z$/$Z_{\odot}$, a lower $F_{esc}$ results
in lower Ly$\alpha$/HeII and higher $\eta_{Ly\alpha}$. In
Fig. ~\ref{lya_filt_a} we show an alternate version of this Figure,
plotted with U scaled by 1/$F_{esc}$ to aid comparison with our other
models.} 
\label{lya_filt}
\end{figure}

\begin{figure}
\includegraphics{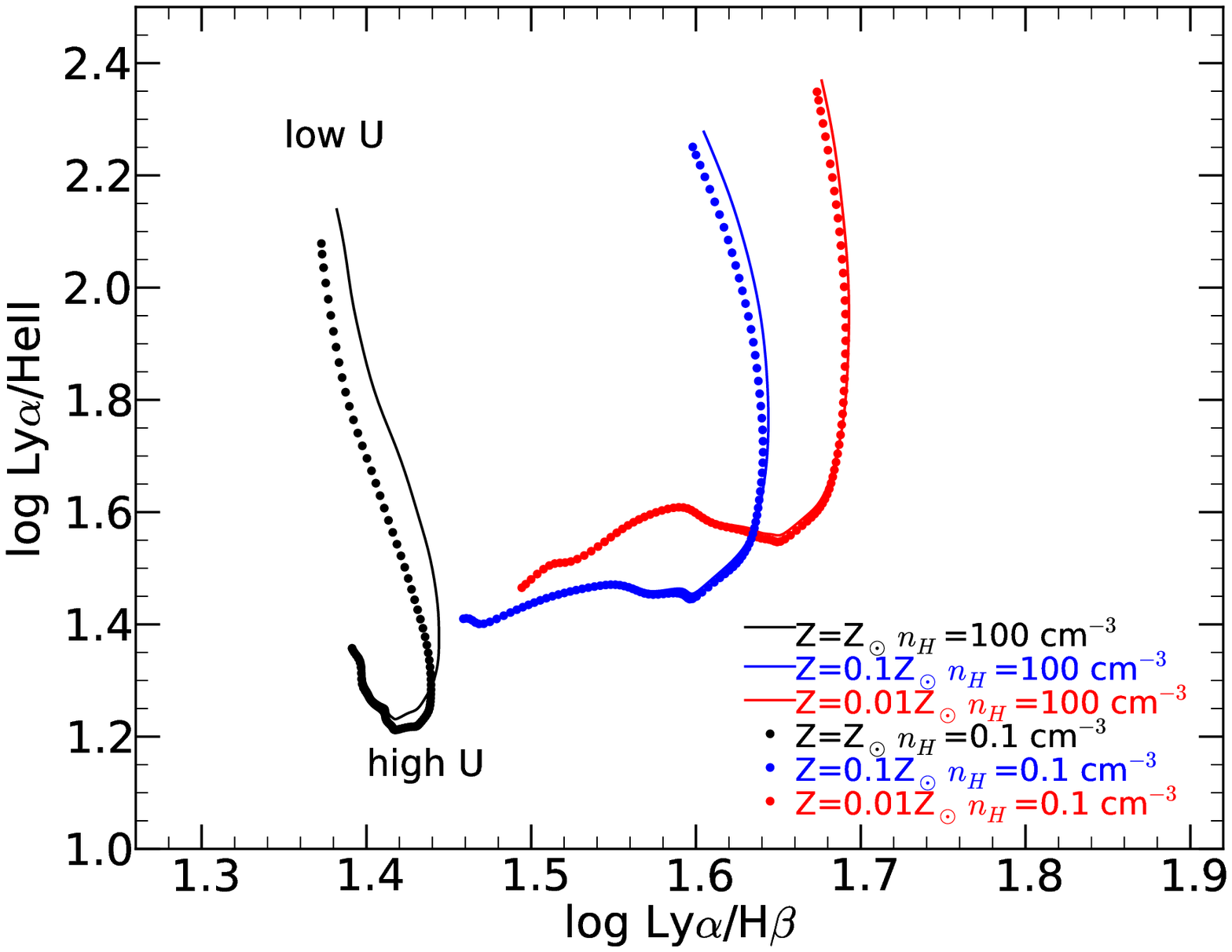}
\includegraphics{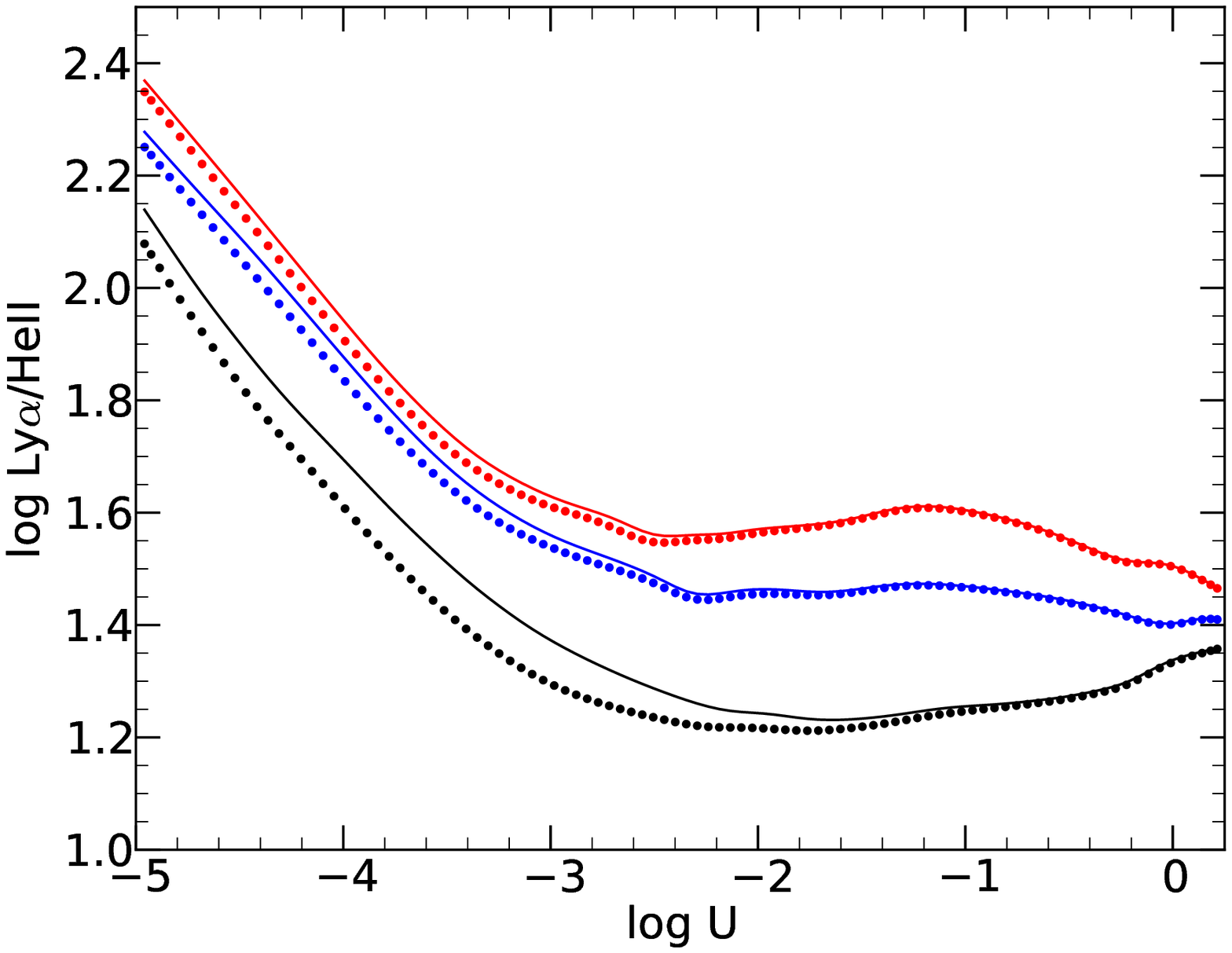}
\includegraphics{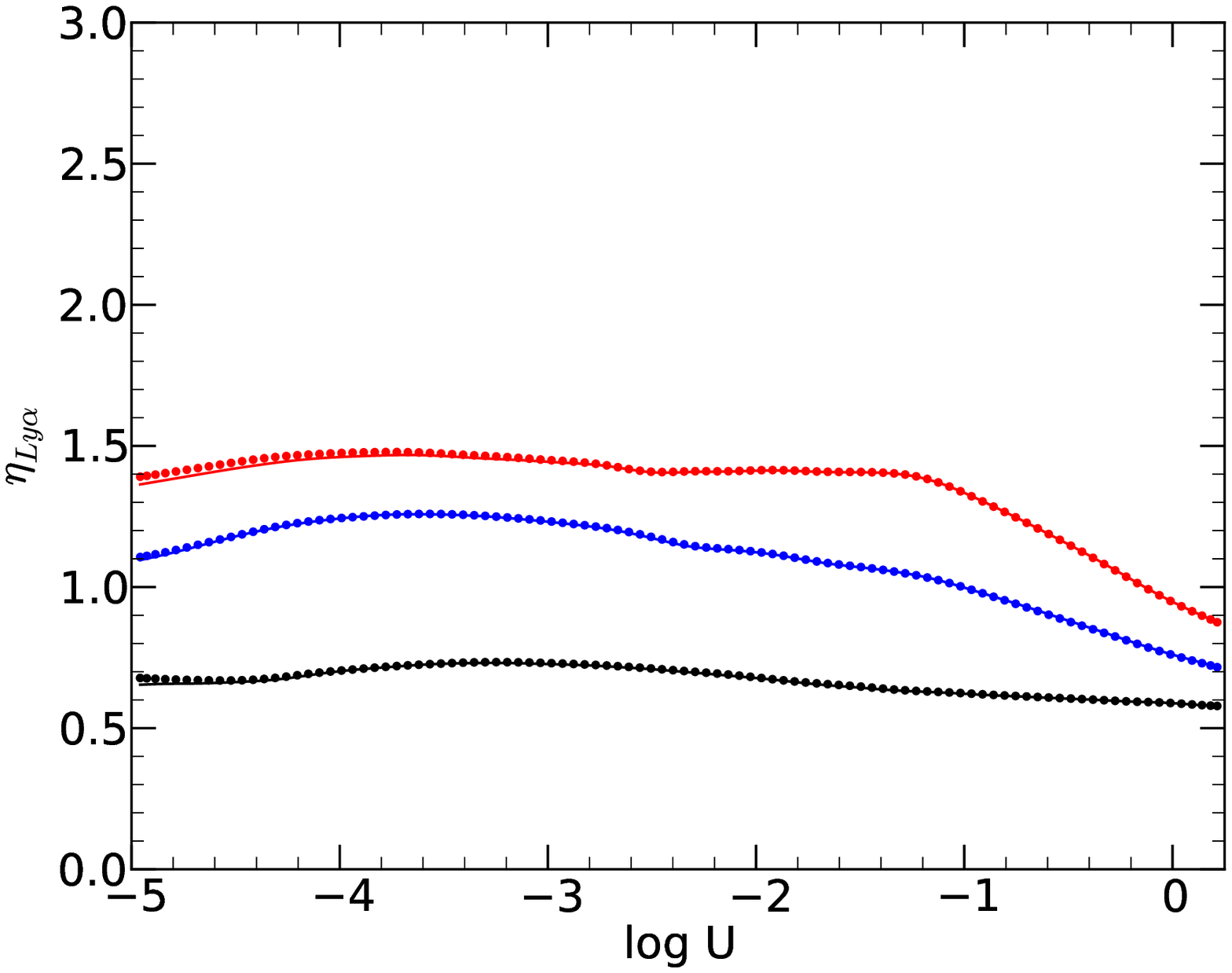}
\vspace{9in}
\caption{The impact of gas density on the observed Ly$\alpha$/HeII,
  Ly$\alpha$/H$\beta$ and $\eta_{Ly\alpha}$ values. The model loci
  cover the range of ionization parameter -5$<$log U$<$0.25.}   
\label{density}
\end{figure}

\subsection{Ionization by Pop III stars}
\label{bb}
We also include a model sequence to simulate a low-metallicity nebula
being photoionized by Pop III (or Pop III-like) stars, with the aim of
understanding to what extent their Ly$\alpha$ flux ratios differ from
those of our AGN models.

Taking into account the effect of He$^+$ opacity, a zero-metallicity
star with an effective temperature of 80~000 K would be equivalent to
a black-body of 67 200 K in terms of the proportion of He$^+$-ionizing
photons (see Holden et al. 2001; Schaerer 2002; Binette et
al. 2003). Thus, we adopt a black-body temperature of 67~200 K in this
model sequence.

We also use a gas phase metallicity of $Z/Z_{\odot}$=0.01, to account
for some slight pollution with metals, and for consistency with the 
low-metallicity AGN powerlaw models against which our Pop III models
will be compared. For easy comparison with our AGN models, our Pop III
models use an ionization-bounded, plane-parallel geometry and $n_H=$100
cm$^{-3}$.

It is not our intention to model in detail the expected line spectrum
of a Pop III-ionized nebula, but to reach a general overview of the
potential for such a nebula to be confused with an AGN-photoionized
nebula when only the strongest UV emission lines are available. 

We stress that this model sequence may be unrealistic in that true Pop
III stars are expected to form from chemically primordial gas, with
the associated HII region being similarly devoid of metals. Our model
sequence, however, does consider a zero-metallicity Pop III star (or
stars), but the gas it photoionizes has already undergone some
slight polution with metals. In a future paper we will present a more
detailed modeling of emission line diagnostics for Pop III HII-regions
and Pop III galaxies.

\section{Results}
Here we describe the results from our model grid using diagnostic
diagrams, which we show in
Figs. ~\ref{lya_balmer}--~\ref{lya_bb}. 

\subsection{AGN models -- line ratios}
\label{results}

Our modeling shows that `extreme' ratios of Ly$\alpha$/HeII can be
readily produced using AGN photoionization. Within our model grid this
flux ratio spans a huge range in values, from log
Ly$\alpha$/HeII$\sim$1.1 to $\sim$4.6, without the need to invoke
physically implausible values for model parameters. Below we describe
the impact of individual parameters on the line ratios, giving
examples to quantify the impact at specific positions within our
parameter space. 

We find that Ly$\alpha$/HeII is extremely high when U is
relatively low, regardless of which electron energy distribution and
gas metallicity are selected, as can be seen in
Fig. ~\ref{lya_balmer}. For example, at $Z/Z_{\odot}=$0.1 and log 
U = -4 we obtain log Ly$\alpha$/HeII$=$1.88. There is
essentially no upper limit to the Ly$\alpha$/HeII flux ratio, with
Ly$\alpha$/HeII $\rightarrow$ $\infty$ as U $\rightarrow$ 0 (log U
$\rightarrow$ $-\infty$). 

We also note that most of the model sequences show at their high-U end
(log U$\ga$-2.5) a region of relatively constant Ly$\alpha$/HeII
values, confirming that this ratio is relatively insensitive to U in
the high-ionization regime (Fig. ~\ref{lya_balmer}; see also
e.g. Villar-Mart\'{i}n et al. 2007; Arrigoni-Battaia et al. 2015b). 

As shown by previous authors, a reduction in gas metallicity results
in enhanced emission of Ly$\alpha$ relative to HeII due
to increased collisional excitation of Ly$\alpha$ resulting from the
increase in electron temperature (e.g. Villar-Mart\'{i}n et
al. 2007). As shown in Fig. ~\ref{lya_balmer}, our models indicate
that this effect can also strongly enhance 
Ly$\alpha$/H$\beta$. On the other hand, H$\alpha$/H$\beta$ and
Ly$\alpha$/H$\alpha$ show relatively small increases with reducing
metallicity. For instance, when log U = -2 and $\alpha$=-1.5, moving from  
$Z/Z_{\odot}=$1.0 to $Z/Z_{\odot}=$0.1 changes log Ly$\alpha$/HeII from
1.25 to 1.46 (+0.21 dex); log Ly$\alpha$/H$\beta$ from 1.43 to 1.59
(+0.16 dex); Ly$\alpha$/H$\alpha$ from 0.96 to 1.10 (+0.14 dex); 
H$\alpha$/H$\beta$ from 0.46 to 0.49 (+0.03 dex). 

The impact of using the $\kappa$-distribution is complex and varies
across the range in parameter space we examine in this work. Generally
speaking, using $\kappa$=20 results in a minor enhancement of
Ly$\alpha$ relative to HeII, H$\beta$ and H$\alpha$ at high
metallicity ($Z/Z_{\odot}\ga$1.0), but when metallicity is low
($Z/Z_{\odot}\la$0.1) the Ly$\alpha$ is marginally reduced relative to
those lines. For instance, at log U = -2, $\alpha=$-1.5 and $Z/Z_{\odot}\ga$1.0,
log Ly$\alpha$/H$\beta$ changes from 1.43 to 1.46 (+0.03 dex) when
adopting $\kappa=$20, but at $Z/Z_{\odot}\ga$0.1 we find that using
$\kappa=$20 reduces log Ly$\alpha$/H$\beta$ from 1.59 to 1.58 (-0.01
dex). In other words, the impact on Ly$\alpha$ of adopting a
  $\kappa$-distribution with $\kappa$=20 is marginal to insignificant, at
  least in the range of parameter space we consider herein. A full
analysis of the impact of using the KD in 
place of the MBD on the emergent spectrum of active galaxies and
high-z Ly$\alpha$ emitters will be presented in a future Paper (Morais
et al. in prep.).

Fig. ~\ref{alpha} shows the impact of using different ionizing spectral indices
($\alpha$). As expected from previous studies (e.g., Humphrey et al. 2008;
Arrigoni et al. 2015b), the Ly$\alpha$/HeII flux ratio is higher
when using a softer ionizing spectrum, because there are relatively
fewer photons able to ionize He$^+$ ($h{\nu}>$54.4 eV). However, a
softer ionizing SED results in a {\it reduction} in the flux of Ly$\alpha$
relative to H$\beta$ and H$\alpha$, and a reduction in
H$\alpha$/H$\beta$, because a softer SED results in lower
electron temperatures, reducing the importance of collisional
excitation effects on these lines. For instance, at
log U = -2 and $Z/Z_{\odot}$, we find that moving from $\alpha=$-1.0 to
$\alpha=$-2.0 changes log Ly$\alpha$/HeII from 1.17 to 1.44 (+0.27
dex), log Ly$\alpha$/H$\beta$ from 1.59 to 1.37 (-0.22 dex), log
Ly$\alpha$/H$\alpha$ from 1.11 to 0.91 (-0.20 dex), log
H$\alpha$/H$\beta$ from 0.48 (3.04) to 0.45 (2.84), and $<T_{OIII}>$ from
14700 K to 9200 K.

In Fig. ~\ref{mirror} we show the effect of adopting the 'front view'
in our adopted geometry. In our models, the observed flux of
Ly$\alpha$ (relative to other lines) is nearly twice that of the `side
view', with HI in the partially ionized / neutral zone
reflecting almost all of the incident Ly$\alpha$ emission, close to
the factor 2.0 (0.3 dex) theoretical maximum effect for a uniformly
flat `back-mirror' with a covering factor of unity. Interestingly,
Fig. ~\ref{mirror} also reveals a degeneracy between metallicity and
cloud viewing angle, with the `front view' mimicking models that use
the `side view' with a lower gas metallicity. Conversely,`rear view'
models (not shown) result in much lower Ly$\alpha$ flux relative to most
other lines, as there is no direct Ly$\alpha$.

As expected, using a filtered ionizing SED results in an enhancement
of Ly$\alpha$/HeII compared to the original, unfiltered SED
(Fig. ~\ref{lya_filt}). This is primarily because the absorbing screen
preferentially absorbs photons in the range 54.4-200 eV, reducing the
relative number of photons that can ionize He$^+$ ($h\nu \ge$
54.4 eV) compared to H (see Fig. ~\ref{filtered}). The
  enhancement in Ly$\alpha$/HeII is also partly due to increased
  collisional excitation of Ly$\alpha$ (see discussion of
  $\eta_{Ly\alpha}$ in Sect. ~\ref{results2}), via two different
  effects: (i) Collisional excitation of Ly$\alpha$ requires the
  presence of neutral H; because it is relatively harder than the
  unabsorbed powerlaw, the filtered SED is less efficient at
  ionizing H, resulting in a higher H neutral fraction and thus a
  higher rate of collisional excitation of Ly$\alpha$; (ii)
  Photoionization by the filtered (harder) SED produces photoelectrons 
  that are on average more energetic, increasing the heating rate and
  temperature of the gas, again leading to a higher rate of
  collisional excitation of Ly$\alpha$. To corroborate the presence of
  these effects, we show in Table ~\ref{table:1} the Ly$\alpha$/HeII
  ratio and its relation to gas metallicity, electron temperature, and
  choice of SED. It is important to note that even though using a
  filtered continuum does not always result in a higher average
  electron temperature, it does increase the Balmer decrement,
  indicating the increased importance of collisional excitation.

The impact on Ly$\alpha$/HeII can be large, even when the filtering screen has a
high escape fraction. For example, using our $F_{esc}=$0.90 SED we
obtain a Ly$\alpha$/HeII ratio that is $\sim$0.2 dex higher than the
unfiltered SED. Even for $F_{esc}$ as high as $\sim$0.97 (not shown
here), this line ratio is still significanly enhanced (+0.04 dex or
10 per cent). At the other end of the parameter range, we obtain log
Ly$\alpha$/HeII $=$ 4.4 using the $F_{esc}=$0.28 SED,
$Z/Z_{\odot}$=1.0 and log U = -5, which rises to 4.6 if we also
include the effect of a `back-mirror'. Using a filtered SED also
results in slight increase in Ly$\alpha$/H$\beta$, of up to $\sim$0.1
dex in the extreme case of our $F_{esc}=$0.28 SED. We also note that
lower ionization lines tend to become stronger relative to
high-ionization lines and HeII (see e.g. Binette et al. 2003). 

We find that gas density has a small, but potentially
  significant effect on the ratios Ly$\alpha$/HeII when log U$\la$ -2
  and $Z/Z_{\odot}$=1.0 (Fig ~\ref{density}). In this parameter range, Ly$\alpha$/HeII is
  up to 0.1 dex higher when using $n_H$=100 cm$^{-3}$ as compared to
  an equivalent $n_H$=0.1 cm$^{-3}$ model. However, at higher values
  of U, or at low gas metallicity ($Z/Z_{\odot}\le$0.1), we find no
  significant difference in the Ly$\alpha$/HeII ratio between our
  $n_H$=0.1 and $n_H$=100 cm$^{-3}$ models.

We note that combining several of the above effects can have a
cumulative enhancement on the Ly$\alpha$/HeII ratio. This is
clearly evident in Figs. ~\ref{lya_balmer}, ~\ref{alpha},
~\ref{mirror} and ~\ref{lya_filt}. For instance, we find that
photoionization at low U (e.g. log U $\la$ -5) by a moderately
filtered continuum (e.g. $F_{esc}$ $\la$ 0.5) can result in very
extreme Ly$\alpha$/HeII flux ratios, with log Ly$\alpha$/HeII
$\ga$4 (Fig. ~\ref{lya_filt}). In another extreme case from our model grid, we
see that low gas metallicity (Z/Z$_{\odot}$=0.01), a relatively soft SED
($\alpha$=-2.0) and low ionization parameter (log U = -5) together
result in log Ly$\alpha$/HeII $=$ 2.88 and log Ly$\alpha$/H$\beta
=$ 1.88 (Fig. ~\ref{alpha}). 

\begin{table*}
\caption{Illustating the impact of low gas metallicity and/or a filtered
ionizing SED on the production of Ly$\alpha$. Columns are as follows:
(1) Gas metallicity normalised to the Solar value; (2) log of the ionization
parameter U; (3) ionizing SED used (including a black-body SED with
T=45,000 K); (4) Average electron temperature
in the H$^+$ zone; (5) hydrogen neutral fraction; (6) ratio of the
column density of ionized hydrogen to the column density of
fully-ionized helium; (7) Balmer decrement; (8) Ly$\alpha$ to HeII
$\lambda$1640 flux ratio; (9) $\eta_{Ly\alpha}$, the ratio of 
Ly$\alpha$ photons emitted to incident ionizing photons. The rows have
been sorted using $\eta_{Ly\alpha}$. As discussed in the main text,
reducing the gas metallicity raises the electron temperature, leading
to increased collisional excitation of Ly$\alpha$ (and of H$\alpha$
relative to H$\beta$), thus increasing $\eta_{Ly\alpha}$. In addition,
adopting a filtered ionizing SED (here we show the $F_{esc}$=0.28
case) has the dual effect of reducing the abundance of He$^{++}$
relative to H$^+$ and increasing the collisional excitation of
Ly$\alpha$, with the latter effect resulting in increased $\eta_{Ly\alpha}$.}
\label{table:1}      
\begin{tabular}{c c c c c c c c c}         
\hline\hline                        
$Z/Z_{\odot}$ & log U & SED & $<T_{H+}>$ (K) & H$^0$ / H$^0$+H$^+$ & H$^+$ /He$^{++}$ & H$\alpha$ / H$\beta$ & Ly$\alpha$ / HeII $\lambda$1640 & $\eta_{Ly\alpha}$ \\
(1) & (2) & (3) & (4) & (5) & (6) & (7) & (8) & (9) \\
\hline                                  
1.0 & -2 & T=45000 K & 6725 & 0.053 & 18705 & 2.80 & 5686 & 0.60 \\
1.0 & -2 & $\alpha$=-1.5 & 11605 & 0.045 & 48.2 & 2.90 & 17.6 & 0.68 \\
1.0 & -2 & $F_{esc}$=0.28 & 13533 & 0.120 & 293.6 & 3.04 & 145.2 & 1.20 \\
0.01 & -2 & $\alpha$=-1.5 & 21384 & 0.054 & 46.4 & 3.13 & 37.0 & 1.40 \\
0.01 & -2 & $F_{esc}$=0.28 & 20667 & 0.148 & 251.3 & 3.29 & 304.9 & 2.80 \\
\hline                                             
\end{tabular}
\end{table*}

\subsection{AGN models -- Ly$\alpha$ to ionizing photon ratio}
\label{results2}
It is also interesting to consider how the luminosity of the
Ly$\alpha$ line varies across our grid of models. To simplify the
comparison between models we use $\eta_{Ly\alpha}$, the ratio of
emergent Ly$\alpha$ photons to incident ionizing
photons. Figs. ~\ref{lya_balmer}, ~\ref{alpha}, ~\ref{mirror} and 
~\ref{lya_filt} each include a panel showing $\eta_{Ly\alpha}$ vs
log U. Our fiducial model sequence using $\alpha$=-1.5 and
$Z$/$Z_{\odot}$=1 (Fig. ~\ref{lya_balmer}) produces values of
$\eta_{Ly\alpha}$ that are close to the value
  $\eta_{Ly\alpha}$=0.66 expected for complete absorption of the
  incident ionizing spectrum and purely recombination emission from
  gas which is optically thick in the Lyman line series (Case B;
  e.g. Gould \& Weinberg et al. 1996). However, we find significant
variation in $\eta_{Ly\alpha}$ across our model grid, which we
  describe below.

As shown in Fig. ~\ref{lya_balmer} (lower right), the gas metallicity can have a
strong effect on $\eta_{Ly\alpha}$, with lower metallicities resulting
in higher values of $\eta_{Ly\alpha}$ (see also Table
~\ref{table:1}). For instance, reducing 
$Z$/$Z_{\odot}$ from 1 to 0.01 increases $\eta_{Ly\alpha}$ by a factor
of $\sim$2. Interestingly, some of our model loci show a substantial
drop in $\eta_{Ly\alpha}$ at higher values of U (log U $\ga$-2),
particularly in model sequences which use low gas metallicity
($Z/Z_{\odot} \le$ 0.1) and a hard ionizing SED
  ($\alpha$$\ge$-1.5; see, e.g., Fig. ~\ref{alpha}). Interestingly,
  when the gas metallicity is sufficiently low ($Z/Z_{\odot} =$ 0.01),
  the rate of collisional excitation\footnote{ 
 In the low density limit and when the Lyman series is optically thick
 (Case B), 
$\eta_{Ly\alpha} \approx \alpha^{eff}_{2p}/\alpha_B \approx 0.67 (T_{H+}/10^{4}K)^{-0.054}$,
 where $\alpha_B$ is the total recombination coefficient and
 $\alpha^{eff}_{2p}$ is the effective recombination coeffient for
 Ly$\alpha$ (see e.g. Binette et al. 1993b). In this regime, values of
 $\eta_{Ly\alpha}$ $\ga$ 0.7 indicate a significant contribution to
 Ly$\alpha$ production due to collisional excitation, while
 $\eta_{Ly\alpha}$ $>$ 1.4 indicates that collisional excitation is the
 dominant channel for Ly$\alpha$ production.} can become the dominant channel
  for the production of Ly$\alpha$ photons, i.e., at $\eta_{Ly\alpha}
  \ga 1.4$.

Fig. ~\ref{lya_balmer} (lower right) also reveals that the choice of 
electron energy distribution can also have a significant impact on
$\eta_{Ly\alpha}$. At high gas metallicity ($Z$/$Z_{\odot}$=1), we
find that using $\kappa$=20 instead of the MBD
  results in a $\sim$10-20 per cent enhancement of $\eta_{Ly\alpha}$
  in the range $-5<$log U$<$-2. However, at low gas metallicity 
(i.e., $Z$/$Z_{\odot}$=0.01), our $\kappa$-distribution models do not show
any significant enhancement in $\eta_{Ly\alpha}$, instead
  showing} a $\sim$10 per cent {\it reduction} in $\eta_{Ly\alpha}$ at
log U$>$-1.4, with no significant difference from the equivalent
  MBD models when log U$<$-1.4.

The hardness of the ionizing radiation also has an impact 
on $\eta_{Ly\alpha}$ (Fig ~\ref{alpha}), with harder ionizing spectra (higher
$\alpha$) resulting in higher values of $\eta_{Ly\alpha}$. As an
example of this, our models with $\alpha$=-1.0 and $Z$/$Z_{\odot}$=1 
yield values of $\eta_{Ly\alpha}$ that are up to $\sim$4 times
higher than produced in our $\alpha$=-2.0 $Z$/$Z_{\odot}$=1
models. This difference widens at lower metallicity: at the
low-metallicity end of our grid ($Z$/$Z_{\odot}$=0.01),
we find that $\eta_{Ly\alpha}$ is a factor of up to $\sim$30 higher
when using $\alpha$=-1.0, compared to $\alpha$=-2.0.

These enhancements of $\eta_{Ly\alpha}$ above the
  expected Case B value are primarily driven by collisional excitation
  of Ly$\alpha$ at the enhanced electron temperatures that result from
  using lower gas metallicity, a harder ionizing spectrum and/or
  $\kappa$-distributed electron energies. The higher
  neutral H fraction produced by the filtered SED can also result in a
  higher rate of collisional excitation of Ly$\alpha$. As a quantitative
  illustration of this effect, when using $\alpha$=-1.5 and log $U$=-2
  we find the average electron temperature in the H$^{+}$ zone is
  T=11605 K when $Z$/$Z_{\odot}$=1, compared to T=21383 K at
  $Z$/$Z_{\odot}$=0.01, resulting in $\eta_{Ly\alpha}$=0.68 and
  $\eta_{Ly\alpha}$=1.40, respectively (see also Table
  ~\ref{table:1}).

Although technically not an enhancement in $\eta_{Ly\alpha}$, the
inclusion of the effect of a `back-mirror`, with the cloud viewed
from the front, does also increase the {\it apparent} (or observed)
$\eta_{Ly\alpha}$ by a factor of $\sim$2 (Fig. ~\ref{mirror}). 

In addition, we find that using a filtered ionizing spectrum
(Fig. ~\ref{lya_filt}) results in an increased
$\eta_{Ly\alpha}$, with more strongly filtered spectra producing higher
values of $\eta_{Ly\alpha}$. This is primarily due to the
  increased importance of collisional excitation under such
  conditions. Note that this effect also results in an increased
  Balmer decrement (see Table ~\ref{table:1}).

We find no significant difference in $\eta_{Ly\alpha}$ between
our $n_H$=0.1 and $n_H$=100 cm$^{-3}$ models (Fig ~\ref{density}).

In summary, we find a substantial variation in $\eta_{Ly\alpha}$, the ratio
of Ly$\alpha$ photons to incident ionizing photons, across our AGN
photoionization model grid. In particular, there is a substantial
deviation from the case B recombination value
  $\eta_{Ly\alpha}$=0.66. In our grid, $\eta_{Ly\alpha}$
ranges from a minimum value of $\eta_{Ly\alpha}$ = 0.58 in the case of
solar metallicity gas illuminated by a powerlaw with $\alpha$=-1.5 at
log U = 0.25, to a maximum of $\eta_{Ly\alpha}$ = 2.7 for a gas cloud
with Z/Z$_{\odot}$=0.01 illuminated at U = -1.6 by our $F_{esc}$ =
0.28 filtered SED. If we also include a neutral 'back mirror', then
the maximum (effective) value in our grid is $\eta_{Ly\alpha}$ = 5.3
(see Fig. ~\ref{lya_filt}).

\subsection{Degeneracy between AGN and black-body models}
\begin{figure*}
\includegraphics{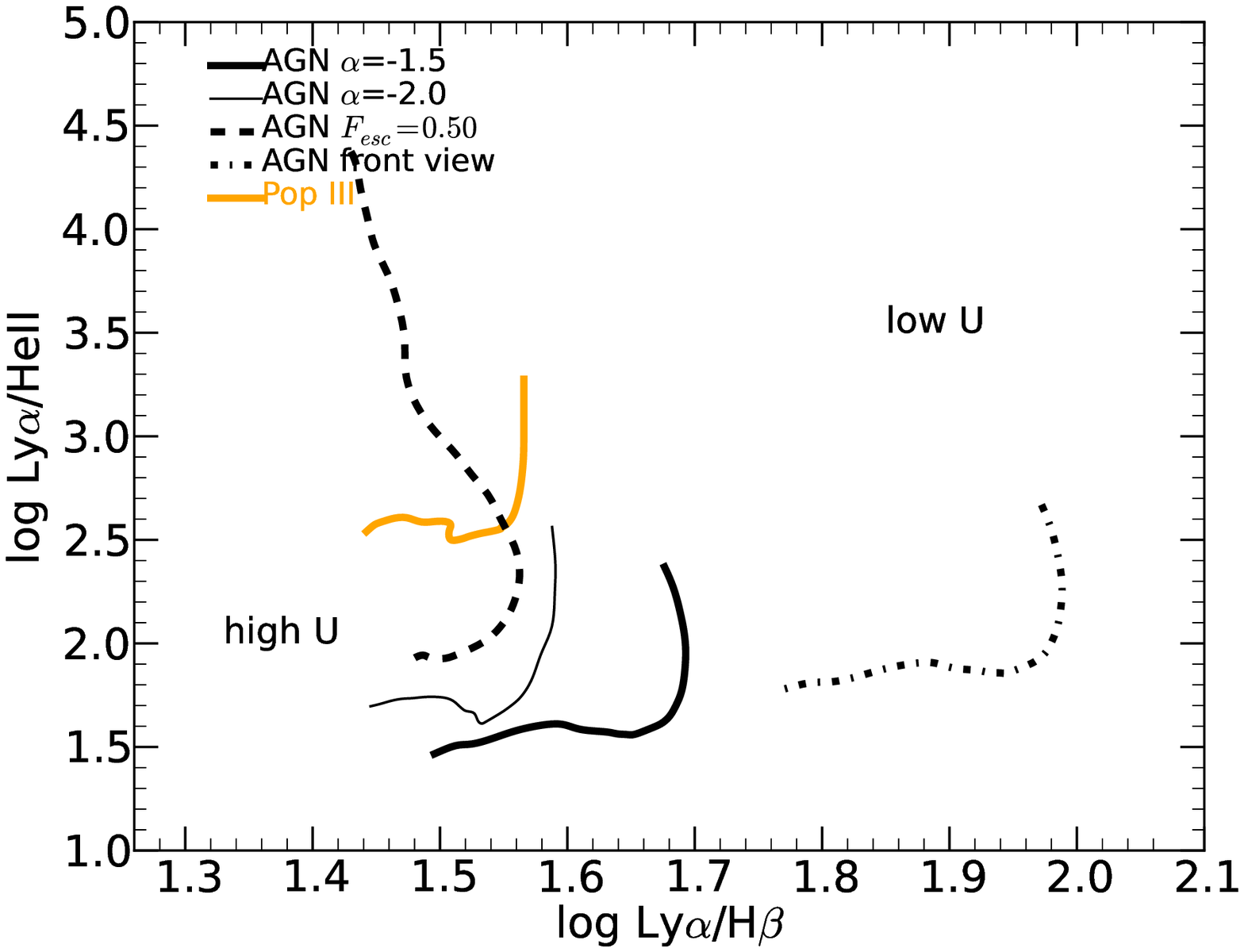}
\includegraphics{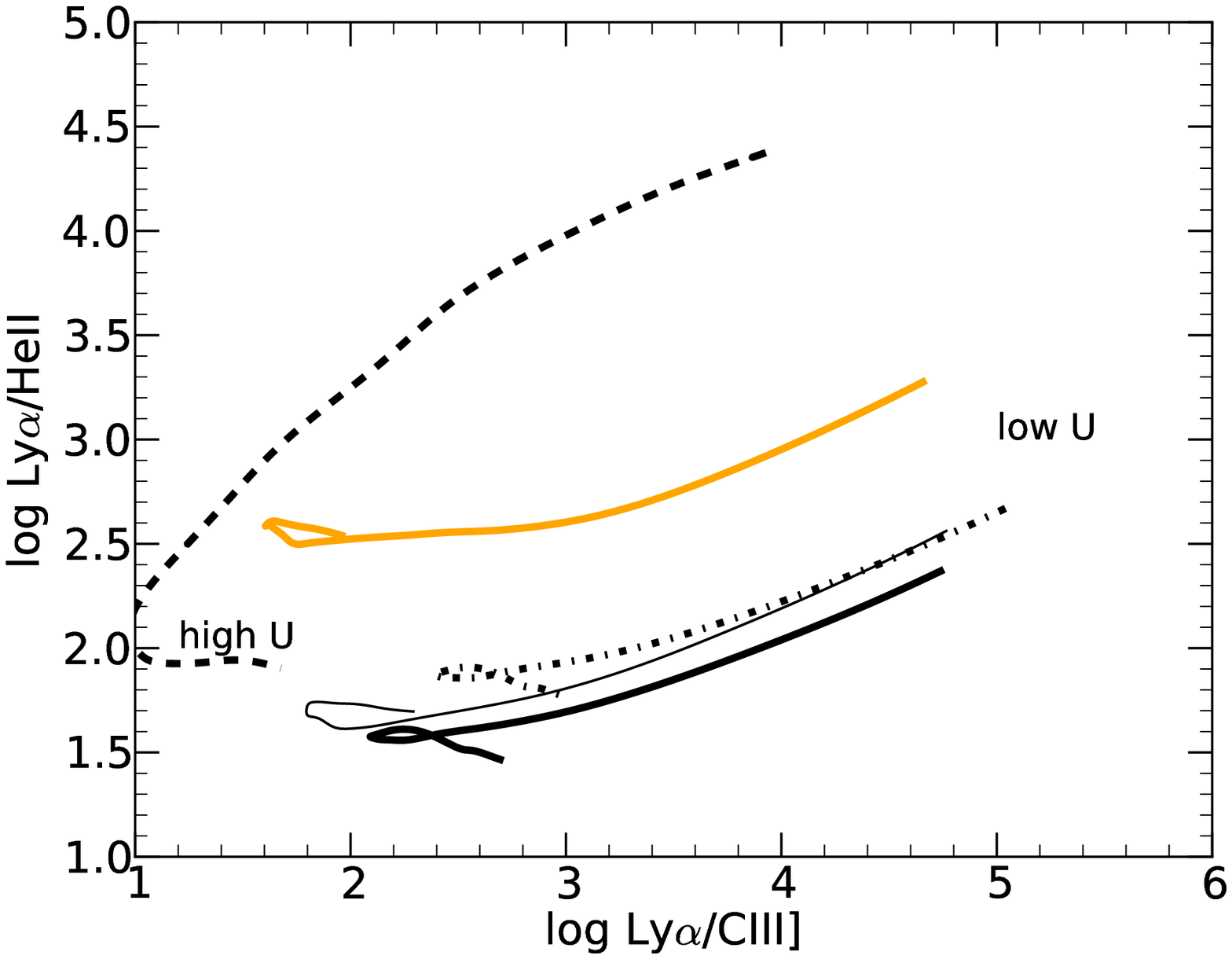}
\includegraphics{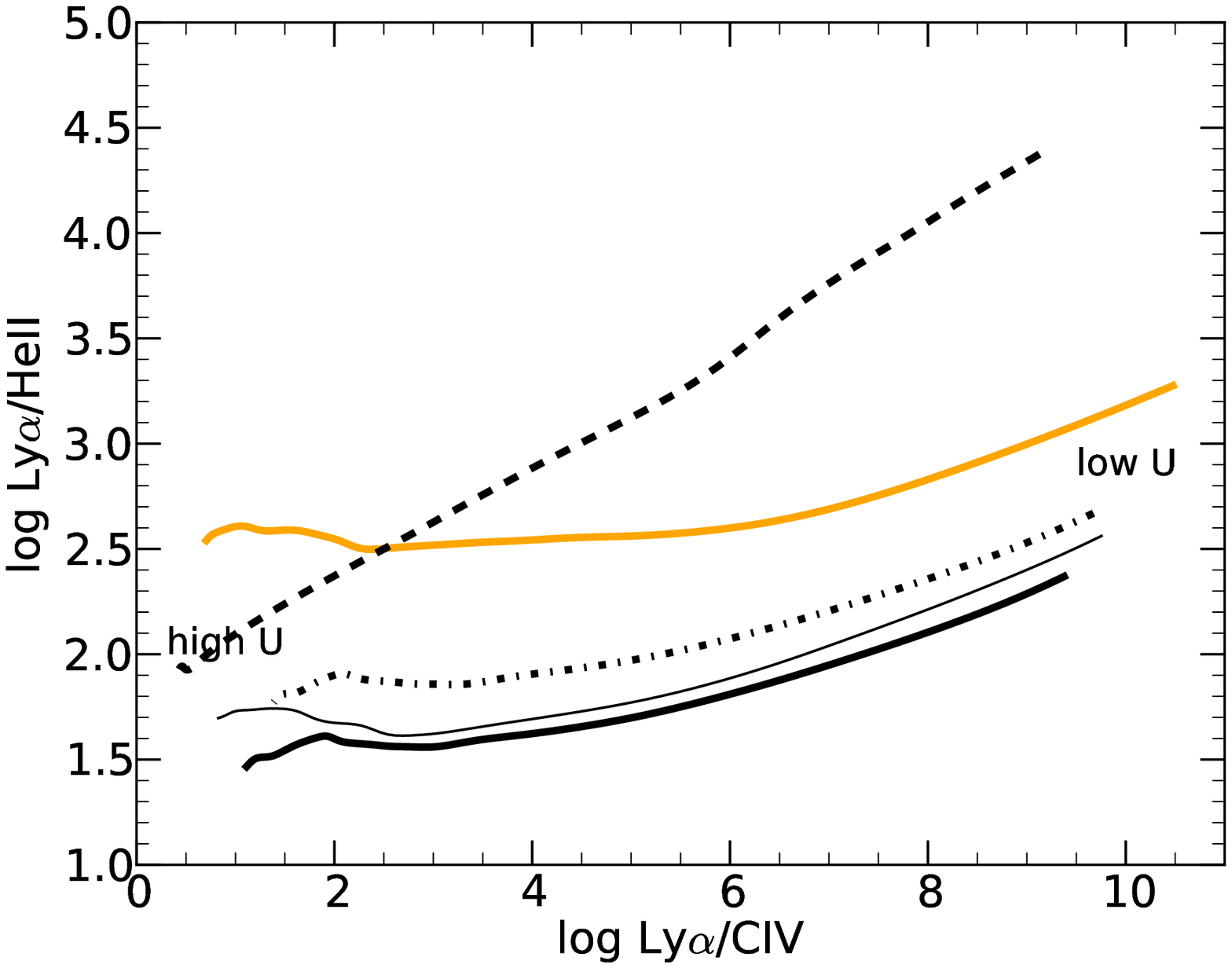}
\includegraphics{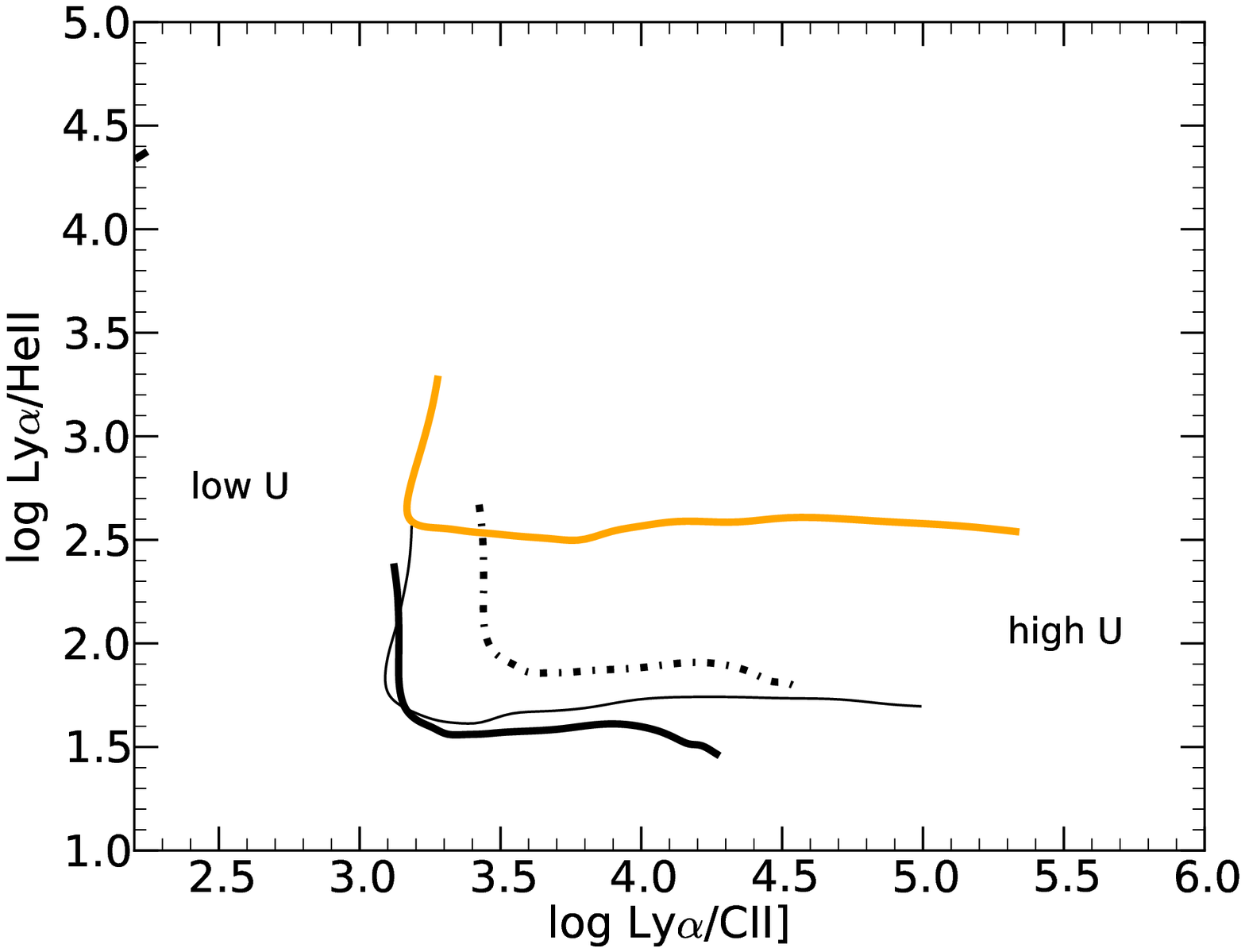}
\vspace{6in}
\caption{Diagnostic diagrams comparing Pop III (heavy orange line) and several AGN
  photoionization models. The AGN model U-sequences shown here are:
  $\alpha$=-1.5 (thick black line); $\alpha$=-2.0 (thin black line);
  $\alpha$=-1.5 with `front view' (`back-mirror') (dot-dashed line); photoionization
  by our $F_{esc}$ filtered AGN continuum (dashed line). All models
  shown in this figure use $Z/Z_{\odot}$=0.01. See main text for
  further details. The model loci cover the full range of ionization
  parameter in our model grid, i.e., -5.55$<$log U$<$0.25.} 
\label{lya_bb}
\end{figure*}

Fig. ~\ref{lya_bb} shows a comparison between our Pop III 
U-sequence and AGN U-sequences with the following parameters: (1) 
$\alpha$=-1.5; (2) $\alpha$=-1.5 with the `front view' of a
photoionized slab with a `back-mirror'; (3) $\alpha$=-2.0; (4) photoionization by an
$\alpha$=-1.5 powerlaw that has been filtered such that
$F_{esc}=$0.50. All the models use $Z/Z_{\odot}$=0.01. 

Comparing models with identical values of U, we find that the Pop III
models give Ly$\alpha$/HeII ratios that are around one order of
magnitude higher than in the powerlaw AGN models. Nonetheless, three of
the AGN sequences do show an overlap in Ly$\alpha$/HeII with the Pop
III sequence. The low-U ends of the $\alpha$=-2.0 and the
$\alpha$=-1.5 `front view' sequences reach up above log 
Ly$\alpha$/HeII$\sim$2.5, while the $F_{esc}$=0.50 filtered continuum
model overlaps with the full range of Ly$\alpha$/HeII values in the
Pop III model sequence. In other words, it is not possible to
distinguish between photoionization by an AGN or by Pop III stars on
the basis of the Ly$\alpha$/HeII ratio alone\footnote{This degeneracy
  disappears when one compares AGN models against 
models for `normal' HII regions, where the ionizing stellar
populations are cooler and result in substantially weaker HeII emission
(see, e.g., Feltre, Charlot \& Gutkin 2016; Sobral et al. 2018)}.

However, the addition of a second line ratio eases the degeneracy
between our model sequences. Of the four diagnostic diagrams
shown in Fig. ~\ref{lya_bb}, the diagram showing Ly$\alpha$/HeII vs
Ly$\alpha$/H$\beta$ yields the cleanest separation of models, with the
AGN sequences falling below and/or to the right of the Pop III model
sequence, because their ionizing SEDs contain a higher proportion of
He$^{+}$ ionizing photons, and/or because the ionized gas has a higher
$T_e$. This diagram also provides a clear diagnostic to
distinguish cases of ionization by a relatively soft source
(e.g., Pop-III stars of an $\alpha$=-2.0 powerlaw) from
AGN-photoionization by an $\alpha$=-1.5 powerlaw with enhanced
Ly$\alpha$ emission due to scattering effects (e.g., our `front view'
models). Using a softer ionizing SED shifts a model upwards, while
scattering-enhanced Ly$\alpha$ shifts a model up and right.

The other three diagnostic diagrams in Fig. ~\ref{lya_bb} show
Ly$\alpha$/HeII vs Ly$\alpha$/CIV $\lambda$1549, Ly$\alpha$/CIII] 
$\lambda\lambda$1907,1909 and Ly$\alpha$/CII] $\lambda$2326. There is
a relatively clean separation between the Pop III sequence and the
power-law AGN models, which run roughly parallel to each other. Only
at their very low-U end do the $\alpha$=-2.0 and `front view' model
sequences cross the locus of Pop III models. 

Interestingly, the filtered AGN continuum sequence crosses the Pop III
sequence in all three of our Ly$\alpha$ to carbon diagrams, indicating
a degeneracy between these two types of model when Ly$\alpha$/HeII is
used together with ratios involving CIV, CIII] and/or CII] (see also
Fosbury et al. 2003; Binette et al. 2003). 

\section{Discussion}
\subsection{On the extreme Ly$\alpha$/HeII ratios around $z>$3 quasars} 
Recently, Borisova et al. (2016) discovered ubiquitous, large-scale
Ly$\alpha$ emitting halos around 19 quasars at
z$\sim$3.5. Interestingly, the authors found extreme Ly$\alpha$/HeII
flux ratios in 11 of the 19 halos, typically with higher
Ly$\alpha$/HeII ratios than in high-redshift radio galaxies. 

A number of earlier observational studies have also detected extremely
high Ly$\alpha$/HeII flux ratios in extended Ly$\alpha$-emitting
regions around distant AGN, with proposed explanations including
enhanced Ly$\alpha$ flux due to the resonance scattering
  reflection of
   Ly$\alpha$ by a large-scale unform halo of neutral or possibly
   matter-bounded gas (see Sect. 3.7), or
recombination emission resulting from photoionization of
  metal-poor gas by the AGN,  (e.g. Villar-Mart\'{i}n et al. 2007;
Humphrey et al. 2013a; Arrigoni Battaia 2015b).

Our modeling has confirmed that scattering effects can indeed lead to
up to a factor of $\sim$2 enhancement of Ly$\alpha$/HeII, under
circumstances that may plausibly be present in the Ly$\alpha$ nebulae of
some high-z AGN (see also, e.g., Villar-Mart\'{i}n, Binette \& Fosbury
1996). However, our models also show that photoionization by an AGN
can produce extreme values of Ly$\alpha$/HeII even without Ly$\alpha$
scattering effects. These include low gas metallicity, a soft or filtered
ionizing SED, or a low ionization parameter. 

\begin{figure*}
\includegraphics{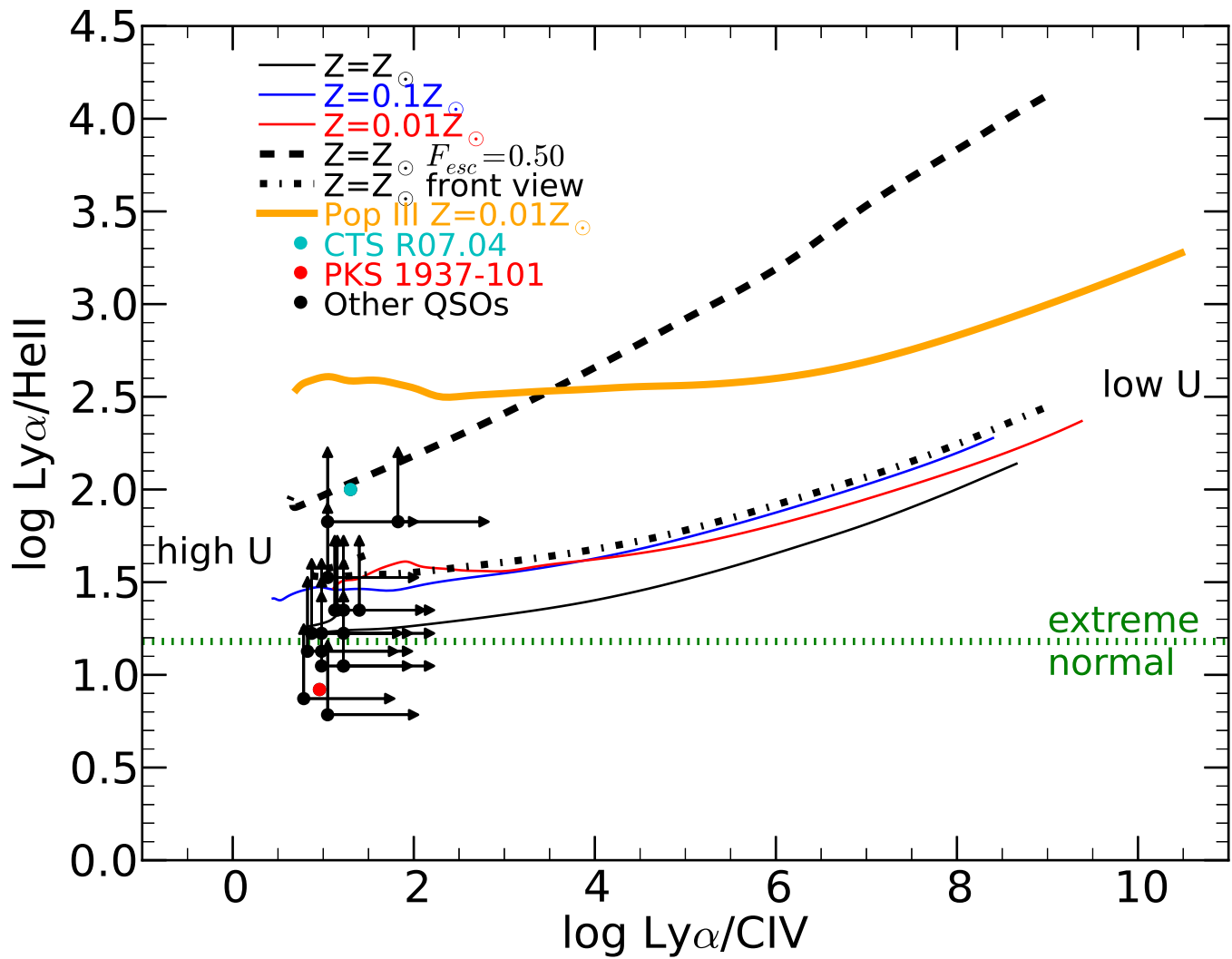}
\vspace{5.85in}
\caption{Plot of Ly$\alpha$/HeII vs. Ly$\alpha$/CIV showing the locus of
  several of our photoionization model sequences, along with the
  measurements or 3$\sigma$ lower limits on these ratios taken from the
  Ly$\alpha$ halos of z$\sim$3.5 quasars of Borisova et
  al. (2016). The model loci cover the full range of ionization 
  parameter in our model grid, i.e., -5$<$log U$<$0.25. The
horizontal green dotted line marks the boundary between what we define
as `extreme' and `normal' Ly$\alpha$/HeII flux ratios.}
\label{lya_heii}
\end{figure*}

In Fig. ~\ref{lya_heii} we show a selection of relevant
photoionization models from our grid, together with the
Ly$\alpha$/HeII flux ratios of Borisova et al. (2016) \footnote{We
  have converted the 2$\sigma$ limits of Borisova et al. (2016) to
  3$\sigma$.}. A horizontal line at log Ly$\alpha$/HeII = 1.18
delineates the `extreme' and `normal'  Ly$\alpha$/HeII regimes defined
in Section 2. 

Interestingly, all of the ionization models shown in Fig. ~\ref{lya_heii}
lie within the `extreme' regime, although at high U the
$Z/Z_{\odot}$=1.0, $\alpha$=-1.5 curve comes close to the
boundary\footnote{Our $Z/Z_{\odot}$=1.0, $\alpha$=-1.0 curve, not
  shown in this Figure, just crosses the boundary and reaches down to
  log Ly$\alpha$/HeII=1.1.}. Thus, we argue that unless log U $\ga$-2.5
{\it and} $Z/Z_{\odot}$$\sim$1, one should expect the intrinsic (emitted)
Ly$\alpha$/HeII ratios of quasar-photoionized halos to
be well within the `extreme' regime. We also suggest that for a
quasar-ionized halo to appear well inside the `normal' regime (log
Ly$\alpha$/HeII $\la$ 1), its Ly$\alpha$ emission would need to be
either strongly absorbed, or else would need to contain a
  substantial population of matter-bounded clouds that are
  sufficiently thin for He to be mostly doubly-ionized.

The majority of the quasars in the sample of Borisova et al. (2016)
have only lower limits on Ly$\alpha$/HeII and
Ly$\alpha$/CIV. Unfortunately, these limits are consistent with
essentially all of the models we have considered, and thus we cannot
place meaningful constraints on properties such as U, $Z$,
$\alpha$, etc. for the gas halos of these particular quasars. 

Only two quasars in the sample of Borisova et al. (2016) have
detections of narrow HeII, and both show large vertical offsets of
$\sim$0.3 dex from our AGN powerlaw models. One of these quasars, PKS
1937-101  at z=3.77, shows log Ly$\alpha$/HeII = 0.92, placing it
within the `normal' Ly$\alpha$/HeII regime in Fig. ~\ref{lya_heii}
(red data point). Its position $\ga$0.2 dex below all of our models
(including our $Z/Z_{\odot}$=1.0, $\alpha$=-1.0 models) suggests that
this quasar halo suffers from strong absorption of Ly$\alpha$ ($N_{HI}
\ga$ 10$^{14}$ cm$^{-2}$), is ionized by an unusually hard SED (I.e.,
$\alpha<$-1.0), or is composed of matter-bounded clouds rather than
the ionization-bounded clouds modeled here. 

The other quasar from this sample with a detection of both HeII and CIV is
CTS R07.04, at z=3.35 (cyan point in Fig. ~\ref{lya_heii}). Its
Ly$\alpha$ halo has log Ly$\alpha$/HeII $\sim$2 and log CIV/HeII
$\sim$5, placing it $\ga$0.4 dex above our powerlaw AGN model loci,
but very close to our $Z/Z_{\odot}$=1.0, $F_{esc}$=0.50 filtered
continuum model locus. We suggest that this quasar halo is ionized by
an unusually soft SED (for an AGN), due to strong filtering of the
quasar's SED by a screen of gas closer to the nucleus, perhaps due to
a wide-angle, AGN-driven outflow closer to the nucleus of the galaxy
itself. Alternatively, in the case of this quasars one could suppose the
presence of a significant contribution from an ionization mechanism
that does not produce strong HeII, such as cooling radiation or
photoionization by hot, young stars.

With the probable exception of CTS R07.04, we find no need for subsolar
gas metallicites, a soft ionizing continuum (including PopIII stars),
or enhanced Ly$\alpha$ by scattering to reproduce the Ly$\alpha$/HeII
flux ratios, although strictly speaking these are not ruled out.

As a consistency check, we can calculate the Ly$\alpha$ luminosity
expected from the implied value of U and the observed size of a
Ly$\alpha$ halo, and then compare against observed values of
Ly$\alpha$ luminosity from Borisova et al. (2016). For photoionization
of gas the expected Ly$\alpha$ luminosity is given by:

\begin{equation}
$$L_{Ly\alpha}~\sim~0.5~U~r^2~n_H~f_c~\eta_{Ly\alpha}~\Omega~($erg
s$^{-1})$$
\end{equation}

\noindent where $f_c$ is the covering factor of the gas as seen by the
ionizing source, $n_H$ (cm$^{-3}$) is the gas density, U is the ionization
parameter, $r$ (cm) is the distance of the cloud from the ionizing
source, $\eta_{Ly\alpha}$ is the ratio of emitted Ly$\alpha$ photons to
incident ionizing photons (see Sect. ~\ref{results2}), and $\Omega$ is
the solid angle (in sr) of the central source covered by the ionized
gas\footnote{The solid angle of an ionization cone with its apex at
  the position of the central ionizing source is given by $\Omega =
  2\pi~(1-cos \beta)$ sr, where $\beta$ is the half opening angle of
  the cone.}. The factor 0.5 arises from the multiplication of the constants
4$\pi$, $h$$\nu_{Ly\alpha}$, $c$, and $1/4\pi$ sr. Due to the expected 
uncertainty in the values of $n_H$, $f_c$ and $\eta_{Ly\alpha}$, we
expect uncertainties in resulting estimates of $L_{Ly\alpha}$ to
be at least a factor of $\sim$10. Throughout this work, we assume
that the ionizing radiation of the AGN is beamed into a bicone covering
a solid angle of $\Omega =$ 3.7 sr, corresponding to a pair of
ionization cones, each with an opening angle of 90$^{\circ}$. 

Taking one of the most `extreme' cases from Borisova et al. (2016), 
Q0042-2627 (z$=$3.3), its Ly$\alpha$/HeII ratio $>$67 (1.8 in log)
would require log U $\la$ -4.2 in our Solar metallicity, $\alpha$$=$-1.5
powerlaw model sequence, corresponding to
  $\eta_{Ly\alpha}\sim$0.66 (determined from
  Fig. ~\ref{lya_balmer}). Although there is  likely to be some radial
  evolution in one or more of $U$, $n_H$, and $f_c$, the exact
  behaviour of these parameters in a real halo is far from 
clear. Thus, we adopt constant but characteristic values for each of
them. Adopting $n_H$$=$100 cm$^{-3}$ and $f_c$$\sim$0.1
estimated from radio-loud, type 2 quasars at z$\ga$2 (e.g., McCarthy
1993; Villar-Mart\'{i}n et al. 2003), assuming $\Omega$ = 3.7 sr,
  and using the maximum observed radius of Ly$\alpha$ emission
$r=$4.9$\times$10$^{23}$ cm (160 kpc) from Borisova et al. (2016), we
estimate $L_{Ly\alpha}$$\la$1.8$\times$10$^{44}$ erg s$^{-1}$ -
  consistent with the observed luminosity
$L_{Ly\alpha}=$1.7$\times$10$^{44}$ erg s$^{-1}$ in Table 2 of
Borisova et al. (2016). The resulting ionizing luminosity of the AGN
would be $Q \la$6$\times$10$^{56}$ s$^{-1}$. Note that the values
  of $L_{Ly\alpha}$ and $Q$ we have derived are both upper limits
  because our estimate of U, from which both are derived, is also an
  upper limit.

Using instead our model sequence with $Z/Z_{\odot}=$0.1 and
$\alpha$=-1.5, we find log U $\la$ -3.8 and $\eta_{Ly\alpha} \la$ 1.25
  from Fig.  ~\ref{lya_balmer}. Leaving $r$, $n_H$, $f_c$ and
$\Omega$ unchanged, we obtain
  $L_{Ly\alpha}$$\la$9$\times$10$^{44}$ erg s$^{-1}$ -- 
  still consistent with the observed luminosity. Using the same
  methodology, we also find a similar level of
consistency between the expected and observed $L_{Ly\alpha}$ 
values for the 10 other quasars from Borisova et al. which show
`extreme' Ly$\alpha$/HeII ratios. 

These calculations serve to illustrate the plausibility of very low
U, possibly coupled with low gas metallicity, to explain the `extreme'
Ly$\alpha$/HeII ratios measured by Borisova et al. (2016). 
  However, we stress that our consistency check is not intended as
  proof of a particular value for any of the parameters in Equation
  (1).

It seems particularly plausible that the large-scale gas halos
of high redshift quasars such as these would have low gas metallicity,
since the central galaxy is expected to be fed by cold streams of
pristine or very low metallicity gas from the cosmic web (e.g. Goerdt
et al. 2015; Vernet et al. 2017). However, observations of the halos
of high-z quasars appear to show such halos are already polluted with
metals (e.g., Humphrey et al. 2013; Prochaska et
al. 2013). 

The detectability of Ly$\alpha$ from extended gas around high-z
  AGN, and the use of this emission for improving our understanding of
  galaxy evolution, continue to be key topics in extragalactic
  astrophysics (e.g., Haiman \& Rees 2001; Villar-Mart\'{i}n et
  al. 2003; Borisova et al. 2016). Our new modeling results have
interesting implications for the detectability of Ly$\alpha$ halos
associated with high-z AGN. Firstly, cold gas around a quasar should
be considerably more luminous in 
Ly$\alpha$, and thus easier to detect, if the gas has a lower
metallicity and/or if it sees a harder ionizing spectrum. We also
suggest that the high detection rate of extended Ly$\alpha$ halos in
quasars at high-z (e.g. Borisova et al. 2016) may be due to this
effect. In addition, lower gas metallicity and a harder ionizing
spectrum could result in apparently larger Ly$\alpha$ halos, as the
correspondingly higher $\eta_{Ly\alpha}$ would make faint, outer
regions more detectable.

\section{Summary}
We have used the modeling code MAPPINGS 1e to explore potential
mechanisms to produce enhanced Ly$\alpha$ relative to HeII and other
emission lines in extended nebulae photoionized by powerful AGN. Our
grid of photoionization models cover a substantial range in ionization
parameter U, the gas metallicity and the shape of the ionizing
continuum, and considers two different electron energy distributions
and cloud viewing angles. 

We are able to produce `extreme' Ly$\alpha$/HeII flux ratios using
parameters appropriate to extended Ly$\alpha$ emitting halos around
high-z quasars. We recover the previously reported
Ly$\alpha$/HeII-enhancing effects of low metallicity (e.g.,
Villar-Mart\'{i}n et al. 2007), low U (e.g. Arrigoni Battaia et
al. 2015a,b), and cloud perspective (Villar-Mart\'{i}n, Binette \&
Fosbury 1996). Our grid reaches much lower values of U
than previous studies by Arrigoni Battaia et al. (2015a,b), and
results in even higher values of Ly$\alpha$/HeII.

As expected from previous studies (e.g., Humphrey et al. 2008;
Arrigoni et al. 2015b), the spectral index of the ionizing powerlaw
affects the Ly$\alpha$/HeII ratio, with a softer spectrum (i.e.,
$\alpha$=-2.0) resulting in higher Ly$\alpha$/HeII values. We also
find that using a pre-filtered ionizing spectrum can result in
extremely high Ly$\alpha$/HeII ratios, with values reaching as high as
log Ly$\alpha$/HeII $=$ 4.4 for heavily filtered continua
($F_{esc}$ = 0.28) and low ionization parameter (log U =
-5.55).

The Ly$\alpha$/HeII-enhancing effects described above can have a
cumulative impact on the Ly$\alpha$/HeII ratio when two or more of the
effects are present in our models. For instance, combining a softer ionizing
continuum, a lower gas metallicity and/or low ionization parameter can
result in a significantly higher Ly$\alpha$/HeII ratio than would have
otherwise be produced using only one of the above. The most
`extreme' model in our grid, which uses log U = -5.55, a heavily
filtered continuum ($F_{esc}$=0.28), low gas metallicity
($Z/Z_{\odot}$=0.01) and a maximal contribution from an HI `back-mirror'
produces log Ly$\alpha$/HeII = 4.6.

In addition to studying the variation of the Ly$\alpha$/HeII flux
ratio, we have also examined the variation in the ratio of emitted
Ly$\alpha$ photons to incident ionizing photons,
$\eta_{Ly\alpha}$. The value of $\eta_{Ly\alpha}$ ranges from 0.58 to
2.7 in our model grid, deviating substantially from the
expected pure-recombination value $\eta_{Ly\alpha}$=0.66 for
optically-thick gas (e.g. Gould \& Weinberg 1996). This variation 
is driven by differences in the electron temperature and H neutral
fraction between models, with higher temperature and higher H neutral
fraction each resulting in higher rates of collisional excitation of 
Ly$\alpha$. In some of our low metallicity models ($Z/Z_{\odot}=0.1$)
we obtain $\eta_{Ly\alpha} >$ 1.4, indicating that collisional
excitation is the main channel of Ly$\alpha$ production. 

In addition, including the effects of an HI `back-mirror' can
  increase the observational $\eta_{Ly\alpha}$ by a factor of up to 2,
  leading to (effective) $\eta_{Ly\alpha}$ values as high as 5.3. Generally
speaking, lower gas metallicities, and/or the use of a harder or filtered ionizing 
SED, results in higher values of $\eta_{Ly\alpha}$. An important implication is
that Ly$\alpha$ halos ought to be easier to detect if they have lower
gas metallicity or are ionized by a harder or filtered ionizing
  SED. In principle, this could lead to selection biases in surveys
  to detect Ly$\alpha$ halos and blobs at high redshift.

Interestingly, we have also found that using $\kappa$-distributed
  electron energies ($\kappa$=20) instead of
  Maxwell-Boltzmann-distributed energies can 
  result in slightly enhanced producion of Ly$\alpha$ photons, with a
  10-20 per cent enhancement in $\eta_{Ly\alpha}$ at high gas
  metallicity ($Z/Z_{\odot}$=1.0 and moderate to low ionization
  parameter (log U $\la$-2). However, at low gas metallicity ($Z/Z_{\odot}\la$0.1) and high
  U (log U $\ga$ -1.4) there is a slight ($\sim$10 per cent) drop in
  $\eta_{Ly\alpha}$. A future study will examine in greater detail the
impact of $\kappa$-distributed electron energies on the emission line
spectrum of the narrow line region or Ly$\alpha$ halo of AGN (Morais
et al., in prep.).

We have also shown that the extreme Ly$\alpha$ and HeII
emission line ratios of the extended Ly$\alpha$ halos of z$\sim$3.5
quasars studied by Borisova et al. (2016) are consistent with
AGN-photoionization of gas with moderate to low metallicity
(e.g. $Z/Z_{\odot}$$\la$0.1) and/or low ionization parameter (e.g. log
U $\la$ -4), without requiring exotic ionization/excitation mechanisms
such as 
PopIII stars or extreme transfer effects. In the case of the most
  `Ly$\alpha$-extreme' quasar from this sample,  CTS R07.04 at
z=3.35, we find that its Ly$\alpha$, HeII and CIV flux ratios are
consistent with gas photoionized by a moderately filtered ionizing SED
($F_{esc} \sim$0.5). We speculate that such filtering may occur in a
wide-angle, AGN-driven outflow nearer the nucleus.

Finally, we find that Ly$\alpha$/HeII alone is
insufficient to discriminate between ionization by Pop III stars and
an AGN. However, we find that using this ratio together with
Ly$\alpha$/H$\beta$ can provide a clean separation between Pop III
and AGN photoionization, even if the gas illuminated by the AGN has
very low gas metallicity ($Z/Z_{\odot}$$\sim$0.01).

\begin{acknowledgements}
We thank the anonymous referee for valuable suggestions that helped
improve this manuscript. AH acknowledges FCT Fellowship
SFRH/BPD/107919/2015; Support from 
European Community Programme (FP7/2007-2013) under grant agreement
No. PIRSES-GA-2013-612701 (SELGIFS); Support from FCT through national
funds (PTDC/FIS-AST/3214/2012 and UID/FIS/04434/2013), and by FEDER
through COMPETE (FCOMP-01-0124-FEDER-029170) and COMPETE2020
(POCI-01-0145-FEDER-007672). AH also acknowledges support from the
FCT-CAPES Transnational Cooperation Project "Parceria Estrat\'egica em
Astrof\'{i}sica Portugal-Brasil". MVM acknowledges support from the
Spanish Ministerio de Econom\'{i}a y Competitividad through the grant
AYA2015-64346-C2-2-P.

\end{acknowledgements}

\begin{appendix}
\section{Alternate version of Fig. ~\ref{lya_filt}}

Here we show an alternate version of Fig. ~\ref{lya_filt}, showing
the same model loci, but plotted as a function of U$_{\ast}$ = U /
$F_{esc}$ to aid comparison with our unfiltered models.

\begin{figure}
\includegraphics{fig_seds_2.ps}
\includegraphics{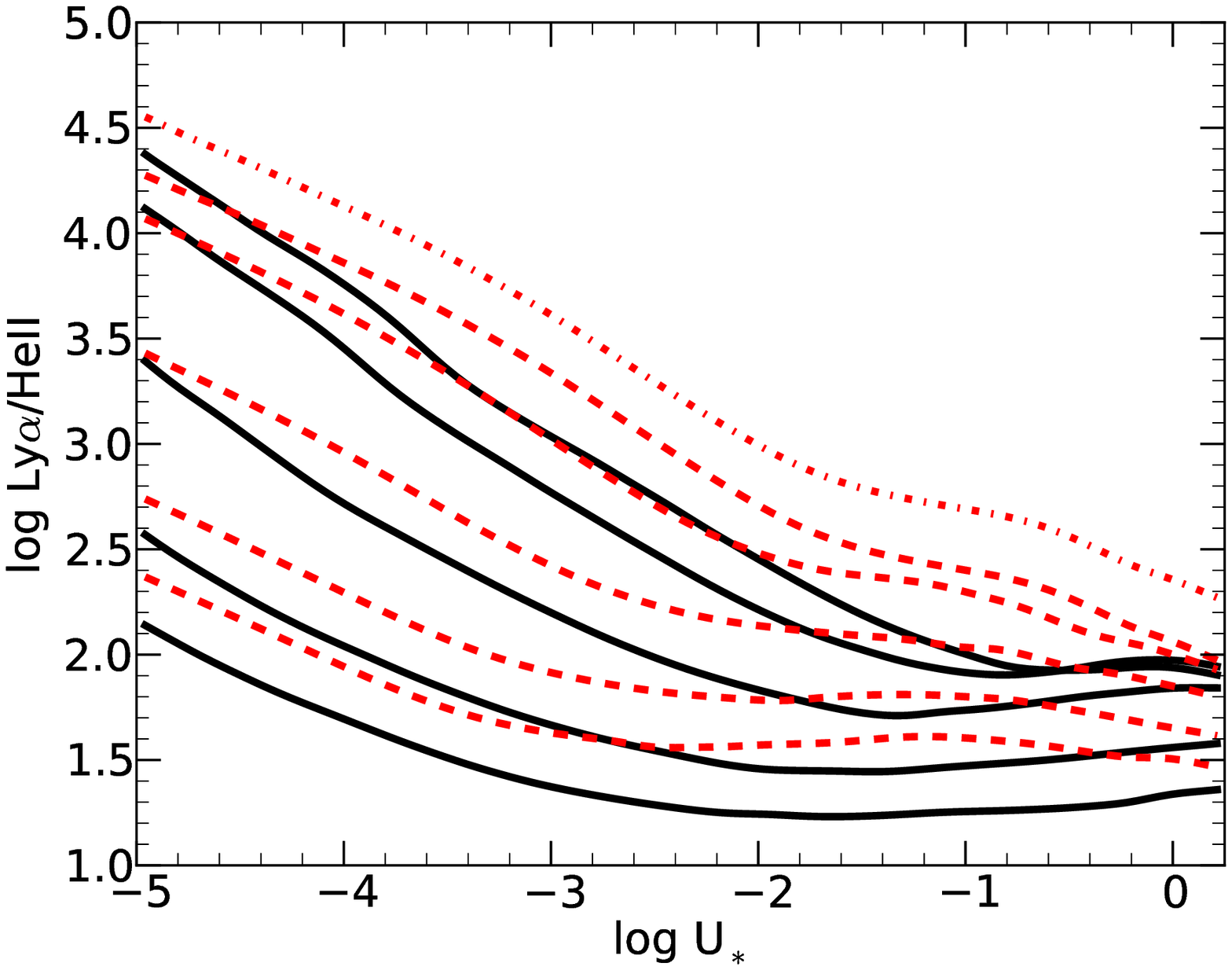}
\includegraphics{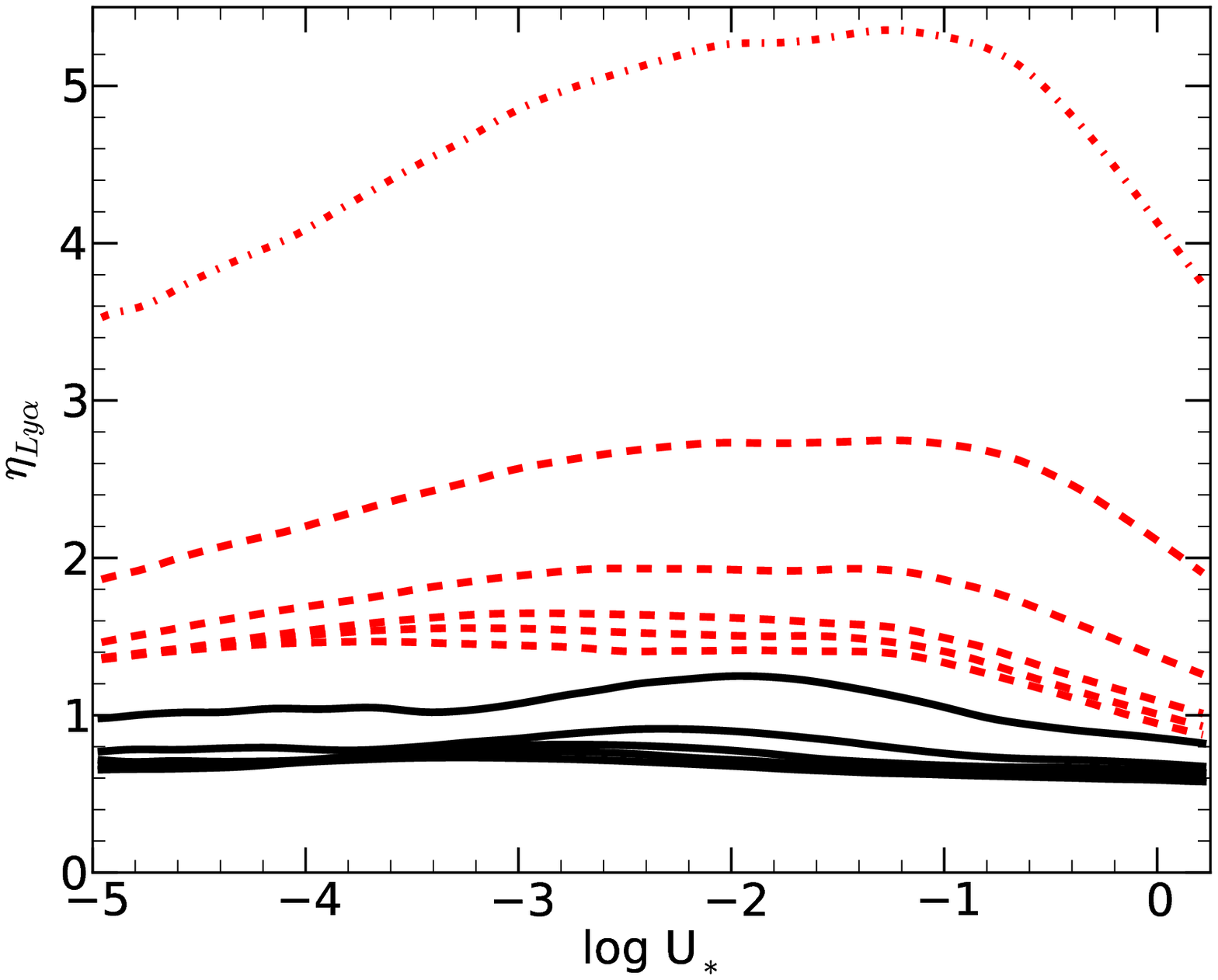}
\vspace{9in}
\caption{The impact of using a filtered ionizing continuum instead of
  a simple powerlaw on the observed values of Ly$\alpha$/HeII,
  Ly$\alpha$/H$\beta$ and $\eta_{Ly\alpha}$. In order that the curves
  be aligned, the models are plotted as a function of log U$_{\ast}$ where
  U$_{\ast}$ = U / $F_{esc}$. In the upper panel, the U$_{\ast}$-sequences using a
  filtered ionizing continuum are labeled with their respective value of
  $F_{esc}$. U$_{\ast}$-sequence loci which use our default powerlaw of
  $\alpha$=-1.5 are labeled '1.0' because the input SED is
  unfiltered. The dot-dashed curve (on the right of the upper panel)
  shows the locus of our sequence in U$_{\ast}$ that uses $F_{esc}$=0.28,
  $Z$/$Z_{\odot}$=0.01, $\alpha$=-1.5 and the `front view', to 
  illustrate the combined effect of low U$_{\ast}$, low gas metallicity, a heavily
  filtered ionizing continuum, and a `back-mirror'. For
each combination of U$_{\ast}$ and $Z$/$Z_{\odot}$, a lower $F_{esc}$ results
in a lower Ly$\alpha$/HeII.}  
\label{lya_filt_a}
\end{figure}

\end{appendix}

\end{document}